\documentclass[12pt,a4paper]{article}
\usepackage[utf8]{inputenc}
\usepackage[T1]{fontenc}

\usepackage{amscd,amsfonts,amsmath,amssymb,amsthm,bbm,bm,latexsym,mathrsfs,mathtools,thmtools}
\usepackage[subnum]{cases}	
\allowdisplaybreaks			

\usepackage{hyperref} 		
\hypersetup{bookmarks=true, unicode=true, urlcolor=blue, linkcolor=blue, filecolor=blue, colorlinks=true, citecolor=black, final=true}

\usepackage{xcolor,subfig,epsfig,graphics,graphicx,rotating}
\usepackage{caption}
\usepackage{enumitem}

\usepackage{longtable,float,array}
\usepackage{setspace}		

\usepackage{threeparttable,booktabs,rotating}

\makeatother
\usepackage{url} 

\usepackage{xspace}
\usepackage{dsfont}

\def\Re {\mathds{R}}

\newcommand{\MATLAB}{\textsc{Matlab}\xspace}
\newcommand{\R}{\textbf{\textsf{R}}\xspace}

\theoremstyle{remark}
\newtheorem{rem}{Remark}
\theoremstyle{plain}

\usepackage{ragged2e,adjustbox}
\usepackage[autostyle,english=american]{csquotes}

\title{Visualizing and comparing distributions with half-disk density strips}

\author{
\hspace*{-20pt}
Carlo Santagiustina\textsuperscript{a}\thanks{e-mail: \href{mailto: carlo.santagiustina@unive.it}{carlo.santagiustina@unive.it}}
\hspace*{20pt}
Matteo Iacopini\textsuperscript{a}\thanks{e-mail: \href{mailto: matteo.iacopini@unive.it}{matteo.iacopini@unive.it}}
 \\ \\
\small \textsuperscript{a}Ca' Foscari University of Venice, Italy
}

\date{\today}

\begin{document}

\maketitle


%
%
%
%

\begin{abstract} 
We propose a user-friendly graphical tool, the half-disk density strip (HDDS), for visualizing and comparing probability density functions. The HDDS exploits color shading for representing a distribution in an intuitive way.

In univariate settings, the half-disk density strip allows to immediately discern the key characteristics of a density, such as symmetry, dispersion, and multi-modality.
In the multivariate settings, we define HDDS tables to generalize the concept of contingency tables. It is an array of half-disk density strips, which compactly displays the univariate marginal and conditional densities of a variable of interest, together with the joint and marginal densities of the conditioning variables.
Moreover, HDDSs are by construction well suited to easily compare pairs of densities.

To highlight the concrete benefits of the proposed methods, we show how to use HDDSs for analyzing income distribution and life-satisfaction, conditionally on continuous and categorical controls, from survey data.

The code for implementing HDDS methods is made available through a dedicated \R package.
\end{abstract}

\noindent%
\vfill

\newpage

\section{Introduction}

\begin{center}
\begin{minipage}[t]{0.85\linewidth}
\it
There is no single statistical tool that is as powerful as a well-chosen graph.
\begin{flushright}
\vspace*{-2.5ex} \cite{chambers2018graphical}
\end{flushright}
\end{minipage}
\end{center}
\vspace*{2ex}

In recent times, methods for the visualization and presentation of data have become increasingly required tools for any quantitative research and business intelligence activity (\cite{Fox2011changing, Shneiderman2014big}).
As initially claimed by \cite{tuckey1977exploratory} and subsequently by \cite{cleveland1985elements} and \cite{tufte2001visual}, graphical methods for exploratory data analysis are extremely useful, especially in the first steps of a statistical study.
Data visualization tools not only affect our capacity to highlight some characteristics of a particular dataset, they also make it possible to rapidly explore the informational content of multivariate datasets, which could otherwise be difficult to convey without any graphical support (\cite{bertin2011graphics}).
As pointed out by \cite{gordon2015statistician}, the quality of a data visualization tool consists in its capacity to be ``interpreted without elaboration''.
According to \cite{vandemeulebroecke2019effective} effective visual communication in quantitative sciences should follow three laws: (i) have a clear purpose; (ii) show the data clearly without distorting its content; (iii) make the message easy to grasp and hence obvious.

In this paper we are concerned with the problem of representing and comparing probability distributions, which is of paramount importance in applied sciences.
The aim is to provide a novel statistically grounded graphical tool that conveys information about univariate and multivariate densities in an intuitive way and allows to make straightforward visual comparisons between distributions, in compliance to the laws (i)-(ii)-(iii).


Uncertainty is a fundamental concept in statistics and probability theory, therefore the development of tools for the visualization of uncertainty is fundamental for the successful communication in any statistical analysis.
Despite the role of uncertainty visualization tools and their impact on research outcome interpretation, which has been recently evidenced by \cite{vandemeulebroecke2019effective} and \cite{seide2020utilizing}, visualization methods often appear to be aesthetic accessories in quantitative researches, and are hence ignored or relegated to a subsidiary role in the literature of statistical methods and applications (\cite{gelman2002let}).
As a result, innovations in visualization methods for statistical data, in particular for multivariate distributions and densities, have been rare and relatively neglected during the last decade.

Standard methods for visualizing probability distributions, such as kernel density plots, histograms, boxplots, and violin plots, are the most widespread tools used in any statistical analysis, although they have several drawbacks.
They may induce a wrong perception of the data, since their clear boundaries can be misinterpreted as precise limits, and thus lead to an automatic ``in or out'' evaluation approach in statistical tests. This is not only contrary to good statistical practice, but also tends to artificially remove any form of uncertainty, contradicting the very purpose of the visualization.
In addition, these plots are often quite hard to read and understand by the general public without a statistical background.

A step forward in this direction has been made by \cite{jackson2008displaying} with the introduction of the \textit{density strip}, a thin horizontal rectangle whose monochrome shade darkness at a point is proportional to the probability density of the variable at that point. This produces a visually appealing display whose continuous gradations match well with the intuition of what uncertainty means.
This approach has been recently popularized by \cite{bowman2019graphics}, who used density strips to visualize uncertainty both of data and in parameter estimation. Furthermore, he exploited interactive plots and animations to convey this information in the 2-dimensional case (e.g., surfaces, spatio-temporal data).
Despite being 2D plots, density strips exploit only the length and give no meaning to the height and the area of the plot.

The aim of this article is to propose novel visualization tools for representing and comparing probability distributions.
We extend the concept of density strips to \textit{half-disk density strips} (HDDSs), which exploits another physical dimension of the plot, the area, to convey an additional layer of information, while retaining an intuitive interpretation. In particular, our approach is particularly well suited for making pairwise comparisons of distributions, and is not affected in its capacity to convey information by the scale of the plot.

The statistical analysis is often concerned with the identification of relationships between variables of interest. In many cases, this quest is motivated by a theoretical result which suggest a certain kind of correlation between two quantities (e.g., economic theory states a positive relationship between disposable income and consumption). However, quite often the data availability comes prior to the development of a strong theory, and this calls for a deep exploratory analysis aimed at discovering potential relationships between the variables in the dataset.
In the context of categorical data, the most popular tool for this purpose is the contingency table, which represents the joint and marginal distribution of two or more categorical variables (e.g., see \cite{fagerland2017statistical}).
A visual inspection of a contingency table provides insights about the presence or absence of a dependence relationship between the variables, which needs to be formally assessed using specific statistical tests.

Consider the problem faced by a decision maker (not necessarily with a strong statistical background) that is interested in visualising the distribution of income, by gender and employment type. Suppose further that data on the same variables are collected for two geographic area and that the decision maker is willing to compare the two associated probability distributions.
Standard contingency tables are not suited for answering these questions.
The main drawback of contingency tables is that they are limited to categorical variables, which greatly restricts their domain of application.
Second, tables for more than two variables may become hard to read, especially for a non-technical user.
Third, since a contingency table is essentially a numerical array reporting the relative frequencies of each event, a comparison between two distributions can be done either by showing two tables or by juxtaposing two numbers in each cell of a unique contingency table. Both solutions are sub-optimal since graphs are better suited than tables to make such comparisons (\cite{gelman2002let}).
Similar drawbacks are carried on to mosaic plots (\cite[][ch.13]{chen2007handbook}), since they provide just a visual representation of contingency tables.

Motivated by these facts, we propose the \textit{HDDS table} as a graphical tool for visualising and comparing conditional and marginal densities. The HDDS table is obtained by arranging HDDSs in an array, which maintains an intuitive interpretation while providing more information than contingency tables.
The advantage of the HDDS table over existing tools is substantial. Using classical visualization methods to report the same information conveyed by the HDDS table requires several plots and numerical arrays (including density plots and contingency tables), with a more complex interpretation. By contrast, the HDDS table provides a compact and intuitive graphical device, which remarkably improves the effectiveness of the communication.

Summarizing, our contribution is twofold.
First, we introduce the half-disk density strip (HDDS) for visualising univariate density functions.
Differently from density strips, which use only a ``virtual'' dimension, the shading, the HDDS also exploits a ``physical'' dimension, the area.
Moreover, the HDDSs can be efficiently stacked to visually compare two distributions, e.g., over time or across countries.
Second, building on this additional layer of information, HDDS tables extend the concept of contingency tables and allow for visualizing and comparing univariate marginal, and conditional distributions.
These methods join a technical background with an intuitive interpretation, making them effective tools especially for communicating with a public that does not have a solid statistical knowledge.

The reminder of the paper is as follows. Section~\ref{sec:HDDS} introduces the HDDS, then Section~\ref{sec:HDDS_table} presents HDDS tables, discusses their interpretation, and provides some examples.
Section~\ref{sec:illustrations} applies the proposed methods to real datasets, and compares the results with alternative standard tools.
Finally, Section~\ref{sec:conclusions} concludes.

The code for implementing the proposed HDDS in \R and \MATLAB is available at:
\begin{center}
\url{https://github.com/matteoiacopini/hdds}
\end{center}

\section{Half-disk density strip}    \label{sec:HDDS}
We define an \textit{half-disk density strip}, or HDDS, as a shaded monochrome sequence of circular sectors, juxtaposed to form a half disk.
The support of the random variable is represented on the semi-circle of the half-disk, from left (lower bound) to right (upper bound), and the the darkness of each circular sector is proportional to the probability density in the interval defined by the corresponding arc.
An HDDS can be seen as a density strip (or DS, \cite{jackson2008displaying}), whose bins are disk sectors rather than rectangles.

For visualizing theoretical densities with unbounded support, we follow the common practice and use a truncation approach.
Any choice of truncation level in the tails, although quite arbitrary, is a practical necessity. However, we remark that since the shading of an HDDS fades gradually to white as the density reduces to zero, the choice of particular bounds for an HDDS do not hamper the results.\footnote{
When the support is unbounded below and above (e.g., real line), an HDDS visualizes the density on an interval that contains the desired central share of the total mass.
Instead, when the support is unbounded from above (e.g., positive real line), the HDDS considers all the values from the lower bound to the desired percentile.
}

\begin{rem}
The HDDS is closely related to the density strip of \cite{jackson2008displaying}.
A distinctive feature of the HDDS which stems directly from its half-disk shape is the invariance to radius scaling.
Given the bounds, the choice of the radius only affects the size of the plot, having no role in the color shading which represents the shape of the distribution.
In practice, in an HDDS one can see the shape of a distribution by looking at the color shading in a small neighborhood of the center or, equivalently, close to the circle border. Moreover, moving from the center towards the circle boundary can be interpreted as a ``zoom-in'' effect: the differences in color shading are visually easier to distinguish, even though no further information is displayed.

Figure~\ref{fig:example_DS} provides a comparison between a DS and an HDDS for representing a mixture of two equally-weighted Normal distributions, $\mathcal{N}(-1.5,1)$ and $\mathcal{N}(1.5,1)$. We set the radius of the HDDS to coincide with the length of the DS and visualise three series of plots, using different value of the radius.
As the length of the radius shrinks, it becomes increasingly hard to distinguish the two modes of the distribution using a DS, while it remains quite easy to spot them in the HDDS.


\begin{figure}[H]
\centering
\setlength{\abovecaptionskip}{0pt}
\begin{tabular}{c c c}
\begin{tabular}{c}
\includegraphics[trim= 10mm 40mm 10mm 10mm,clip,height= 4.4cm, width= 7.0cm]{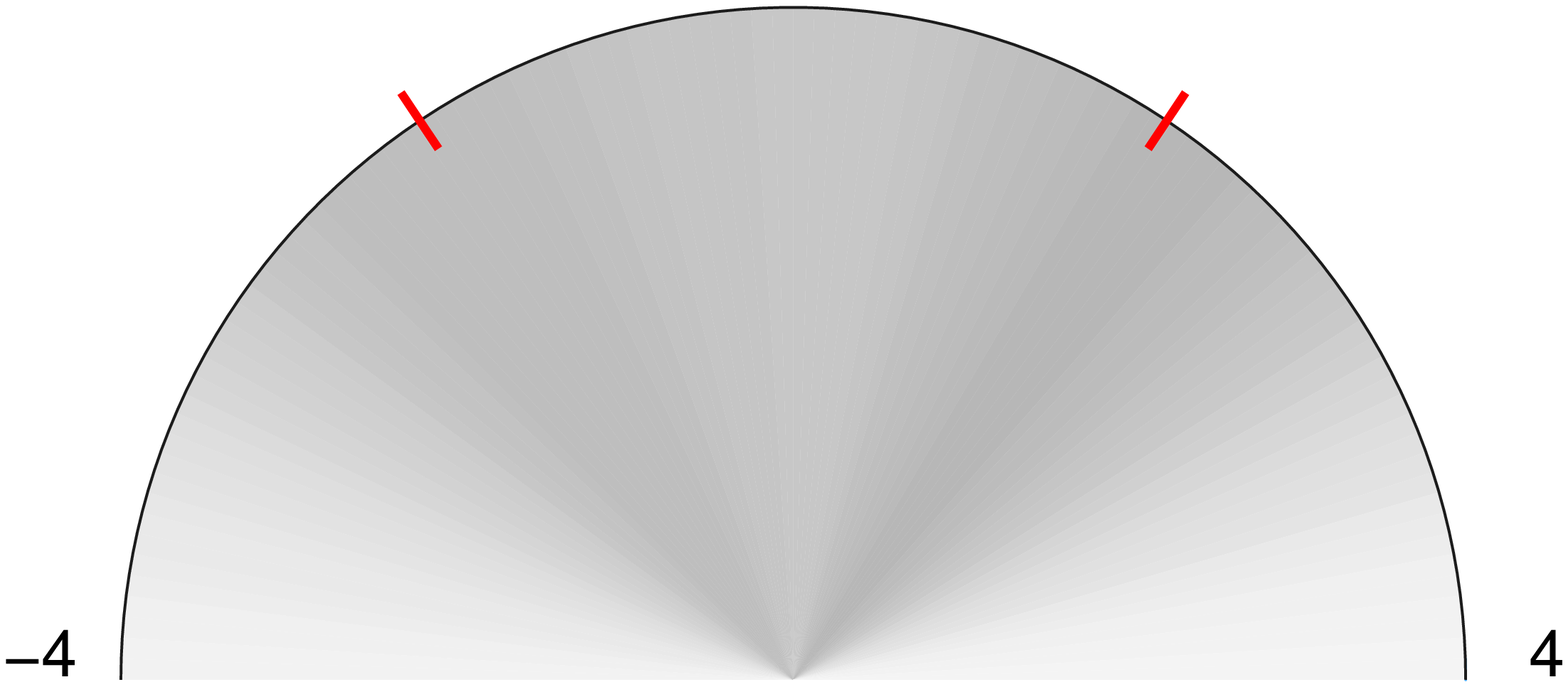} \\[-20pt]
\includegraphics[trim= 5mm 10mm -3mm 20mm,clip,height= 2.0cm, width= 7.0cm]{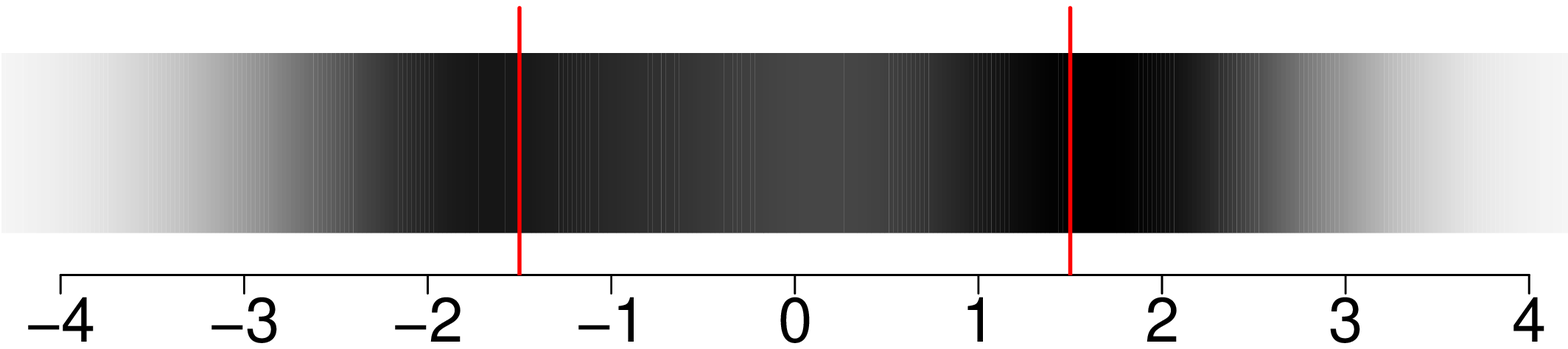}
\end{tabular} &
\begin{tabular}{c}
\includegraphics[trim= 10mm 40mm 10mm 10mm,clip,height= 2.2cm, width= 3.5cm]{example_HDDS_mixture_ticks.eps} \\[-15pt]
\includegraphics[trim= 5mm 10mm -3mm 20mm,clip,height= 1.0cm, width= 3.5cm]{example_DS_mixture_ticks.eps}
\end{tabular} &
\begin{tabular}{c}
\includegraphics[trim= 10mm 40mm 10mm 10mm,clip,height= 1.1cm, width= 1.75cm]{example_HDDS_mixture_ticks.eps} \\[-10pt]
\includegraphics[trim= 5mm 10mm -3mm 20mm,clip,height= 0.5cm, width= 1.75cm]{example_DS_mixture_ticks.eps}
\end{tabular}
\end{tabular}
\caption{Comparison of HDDS (top) and DS (bottom) using three different lengths of the radius of the HDDS (one per each column) for the distribution of 1,600 draws from a mixture of normal $x \sim 0.5 \cdot \mathcal{N}(-1.5,1) + 0.5 \cdot \mathcal{N}(1.5,1)$.
Red ticks correspond to the two modes, $-1.5$ and $1.5$.}
\label{fig:example_DS}
\end{figure}

\end{rem}

\subsection{Color shading}
The darkness of the shading of the HDDS at $x$ is defined as proportional to the value of the density $f(x)$ at that a point.
Let $p = f(x)/\kappa$, where $\kappa = \int f(x) dx$, then the HCL color at $x$ is
\begin{equation*}
p \cdot (cH, cC, cL) + (1-p) \cdot (wH, wC, wL),
\end{equation*}
where $(cH, cC, cL)$ is a dark color chosen for the maximum density and $(wH, wC, wL)$ is white.
The human perception of brightness is nonlinear (e.g., see \cite{poynton2012digital}) and follows an approximate power function, with greater sensitivity to relative differences between darker tones than between lighter tones.
The ``gamma correction'' accounts for this issue and is defined, in the simplest case, by the power-law $V_{out} = V_{in}^{\gamma}$, where $\gamma >0$ and $V_{in}$, $V_{out}$ denote the input and output (real) values encoding the color.
Accordingly, in defining the color shading of an HDDS it is possible to replace $p$ with the monotone transformation $p^\gamma$, which results in darker tails ($\gamma < 1$) or brighter peaks ($\gamma > 1$).

\begin{rem}
For visualizing a distribution, \cite{jackson2008displaying} uses a normalization such that the darkness of the shading of the density strip is maximum at the mode.
This choice is convenient when a single density is displayed by a DS, but it may lead to misleading conclusions when multiple DSs are compared.
For example, consider the DSs of a uniform distribution on $(-1,1)$ and of a Normal distribution $\mathcal{N}(0,0.4^2)$. Using the normalization of \cite{jackson2008displaying}, the DS of the former results in a black strip between $0$ and $1$, while the DS of the Normal has a black shade around $0$, which gradually decays to white on the tails. However, the intensities of the black shade of the (whole) uniform DS and of the Normal DS around 0 are the same, despite the values of the corresponding probability densities are different.
Therefore, when comparing the two DSs, one may have the impression that the total mass of the uniform is greater than that of the Normal, which is not correct.
See Figure~\ref{fig:DS_vs_HDDS}.


\begin{figure}[H]
\centering
\scriptsize
\hspace{-3.5ex}
\setlength{\tabcolsep}{5pt}
\setlength{\abovecaptionskip}{0pt}
\begin{tabular}{c c c}
\begin{rotate}{90} \hspace{12pt} $\mathcal{U}(-1,1)$ \end{rotate} &
\includegraphics[trim=12mm 0mm 12mm 0mm,clip,height= 2.0cm, width=6.0cm]{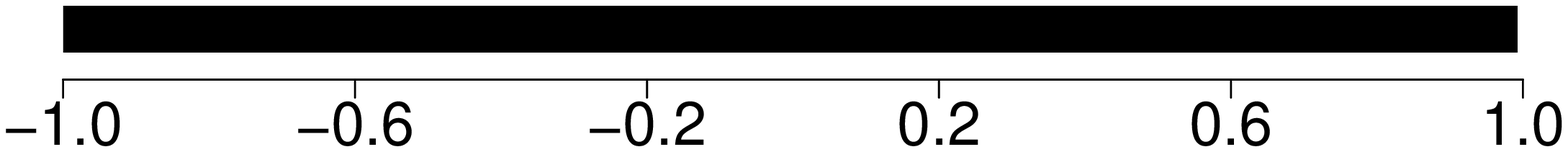} &
\includegraphics[trim=20mm 0mm 20mm 8mm,clip,height= 2.0cm, width=4.2cm]{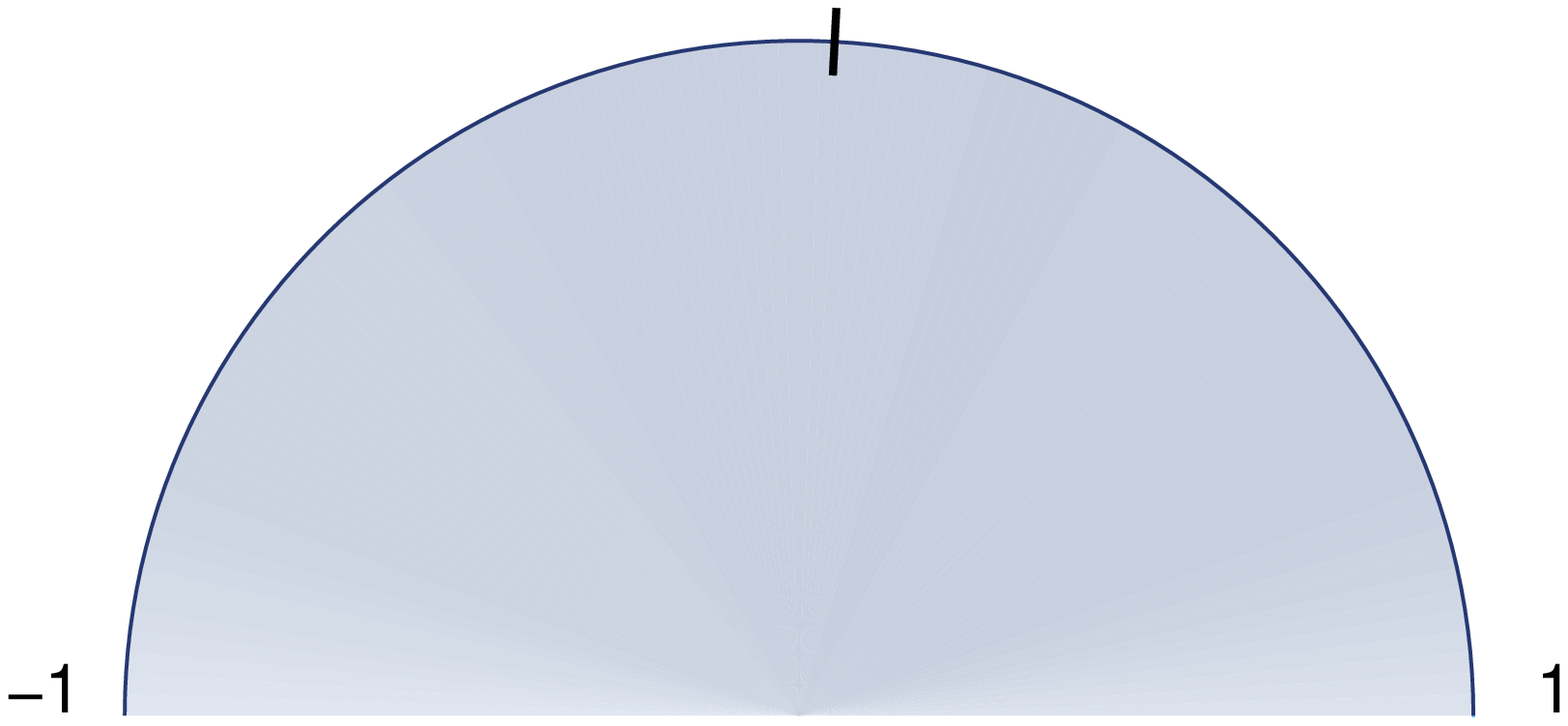} \\[-5pt]
\begin{rotate}{90} \hspace{9pt} $\mathcal{N}(0,0.4^2)$ \end{rotate} &
\includegraphics[trim=12mm 0mm 12mm 0mm,clip,height= 2.0cm, width=6.0cm]{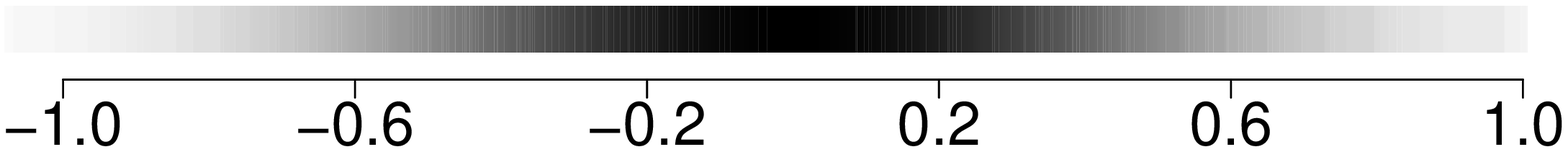} &
\includegraphics[trim=20mm 0mm 20mm 8mm,clip,height= 2.0cm, width=4.2cm]{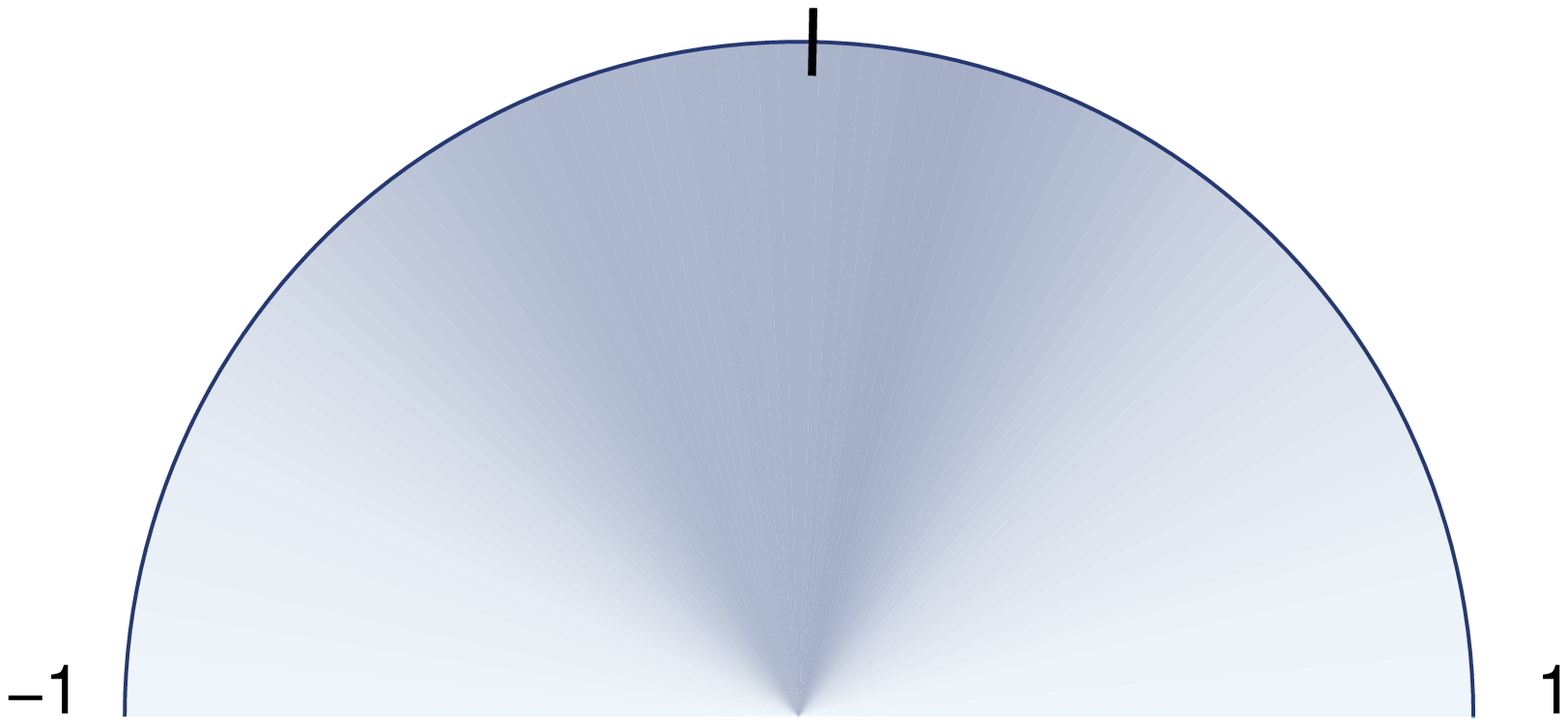}
\end{tabular}
\caption{Comparison of DSs and HDDSs for a uniform $\mathcal{U}(-1,1)$ (top row) and a Normal $\mathcal{N}(0,0.4^2)$ (bottom row).}
\label{fig:DS_vs_HDDS}
\end{figure}

We propose an alternative normalization to solve this issue, by keeping fixed the total amount of color in the graph, and setting the darkness of the HDDS to be proportional to the value of the distribution at a given point.
This normalization is consistent both within and between different HDDSs, thus allowing for correct comparisons of distributions, while DSs are only consistent within themselves.
\end{rem}

The use of colours in plots is an effective mean to convey different types of information. However, colour palettes are not equivalent, since colour selection affects the perception of changes across categories or along continuous scales.
Standard flashy RGB palettes despite attracting the attention of the viewer, may make it hard for him to hold the attention on a plot, since saturated colors can be distracting and produce after-image effects (\cite{ihaka2003colour}).
To deal with this issue, we follow \cite{zeileis2009escaping} and adopt HCL color palettes.
These are cylindrical color space models designed to accord with human perception of color, and can be obtained as nonlinear transformation of RGB triplets.

\subsection{Examples}
The HDDS can provide information about a distribution not only through color shading: for example, the addition of marks on top of it or on its border is straightforward.
In particular, we visualize the median as a thin black line on the border of the HDDS, and explore the possibility of displaying the data points as dots superimposed to the shading.


\begin{figure}[H]  
\centering
\setlength{\tabcolsep}{5pt}
\setlength{\abovecaptionskip}{0pt}
\renewcommand*{\arraystretch}{0.01}
\hspace*{-3.5ex}
\begin{tabular}{c c c c}
\includegraphics[trim= 20mm 10mm 20mm 10mm,clip, height=2.0cm, width=5.0cm]{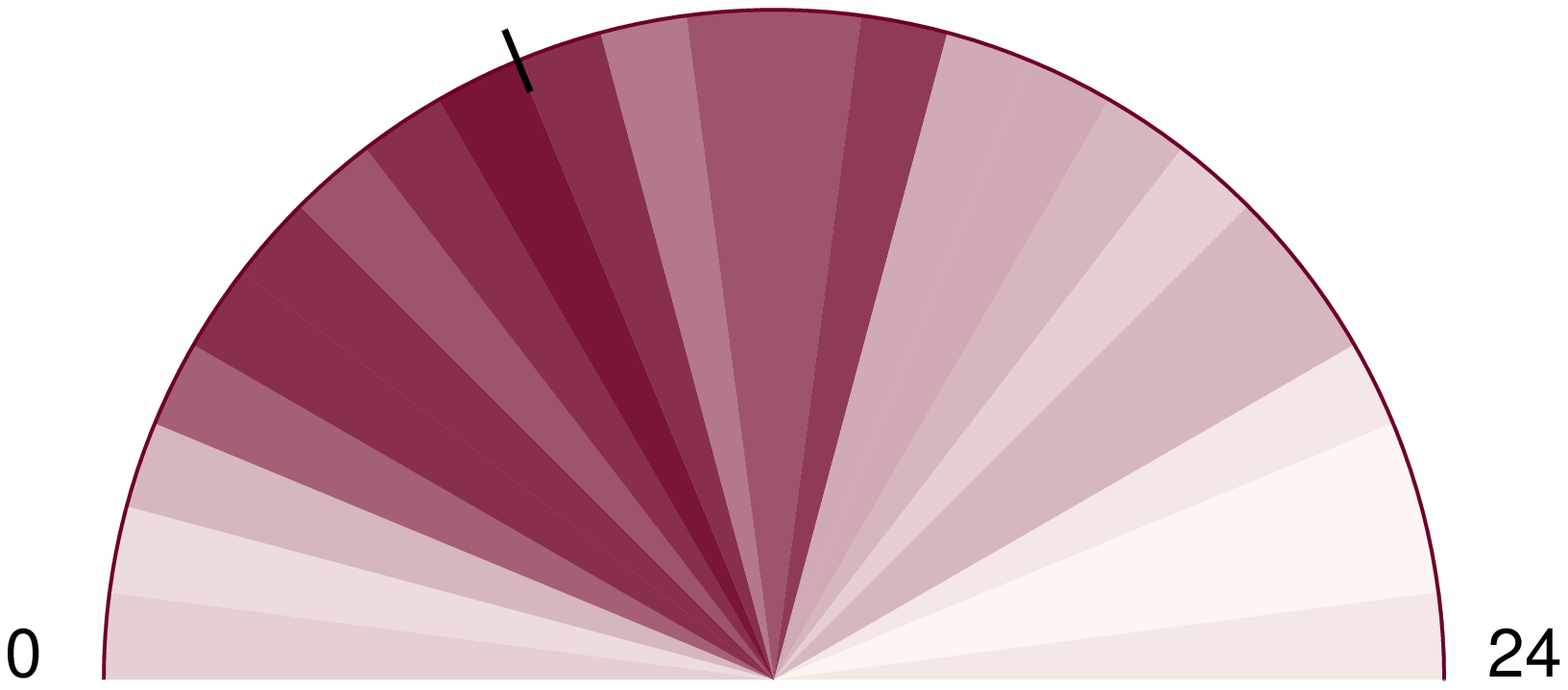} &
\includegraphics[trim= 20mm 10mm 20mm 10mm,clip, height=2.0cm, width=5.0cm]{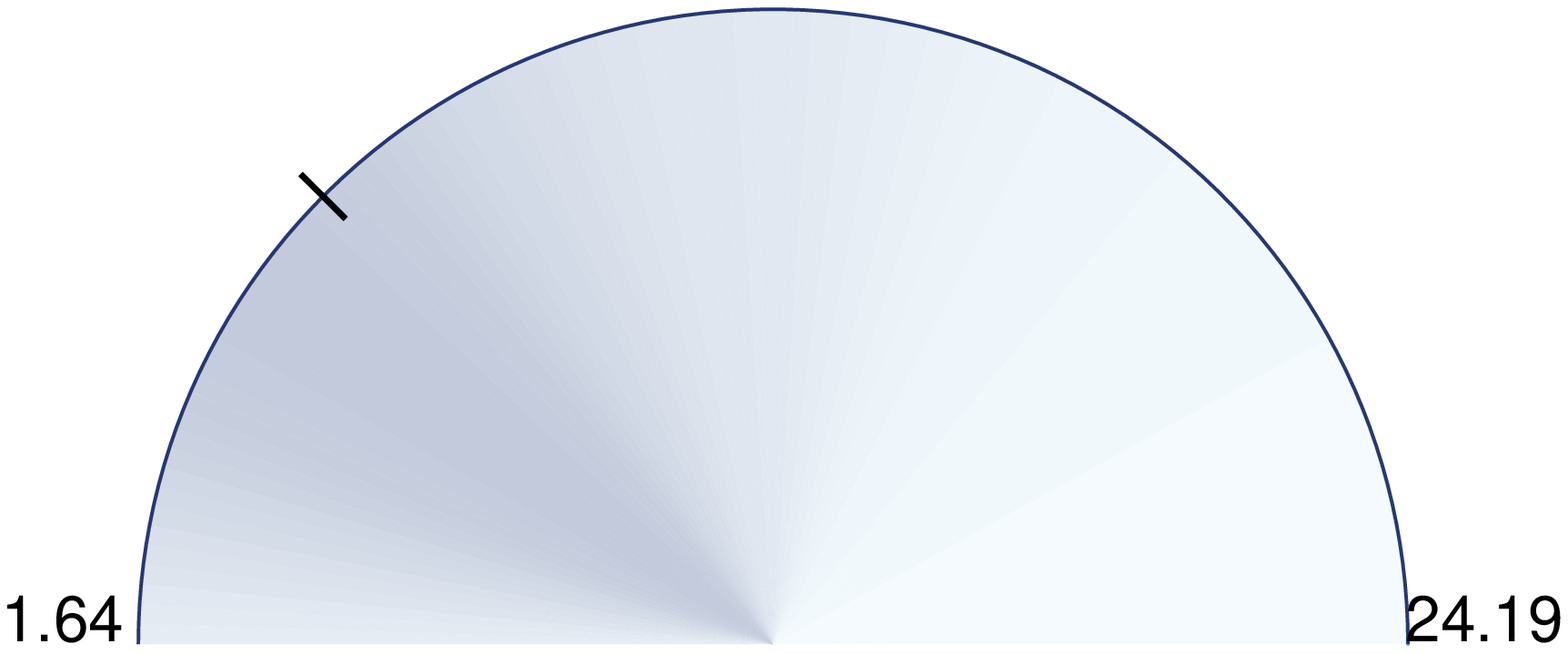} \\
\includegraphics[trim= 20mm 10mm 20mm 10mm,clip, height=2.0cm, width=5.2cm]{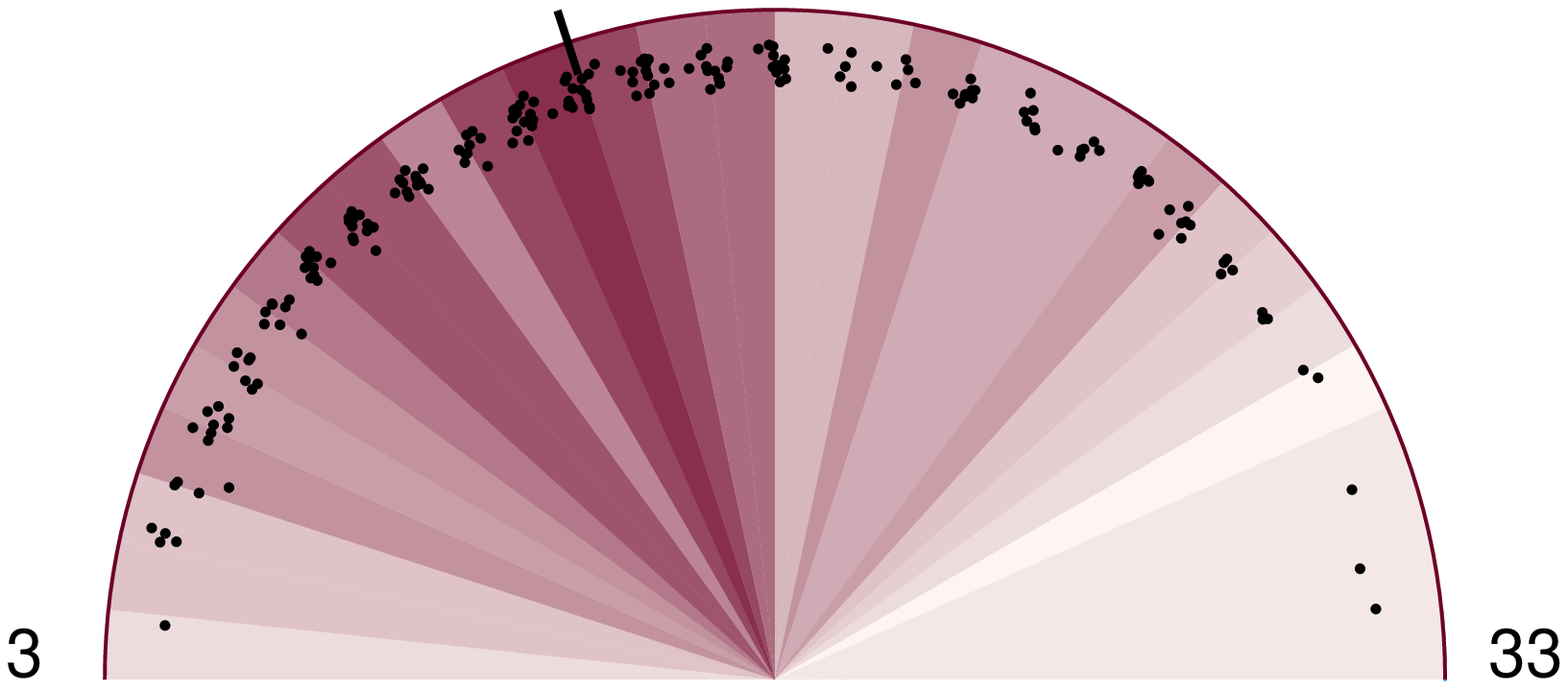} &
\includegraphics[trim= 20mm 10mm 20mm 10mm,clip, height=2.0cm, width=5.2cm]{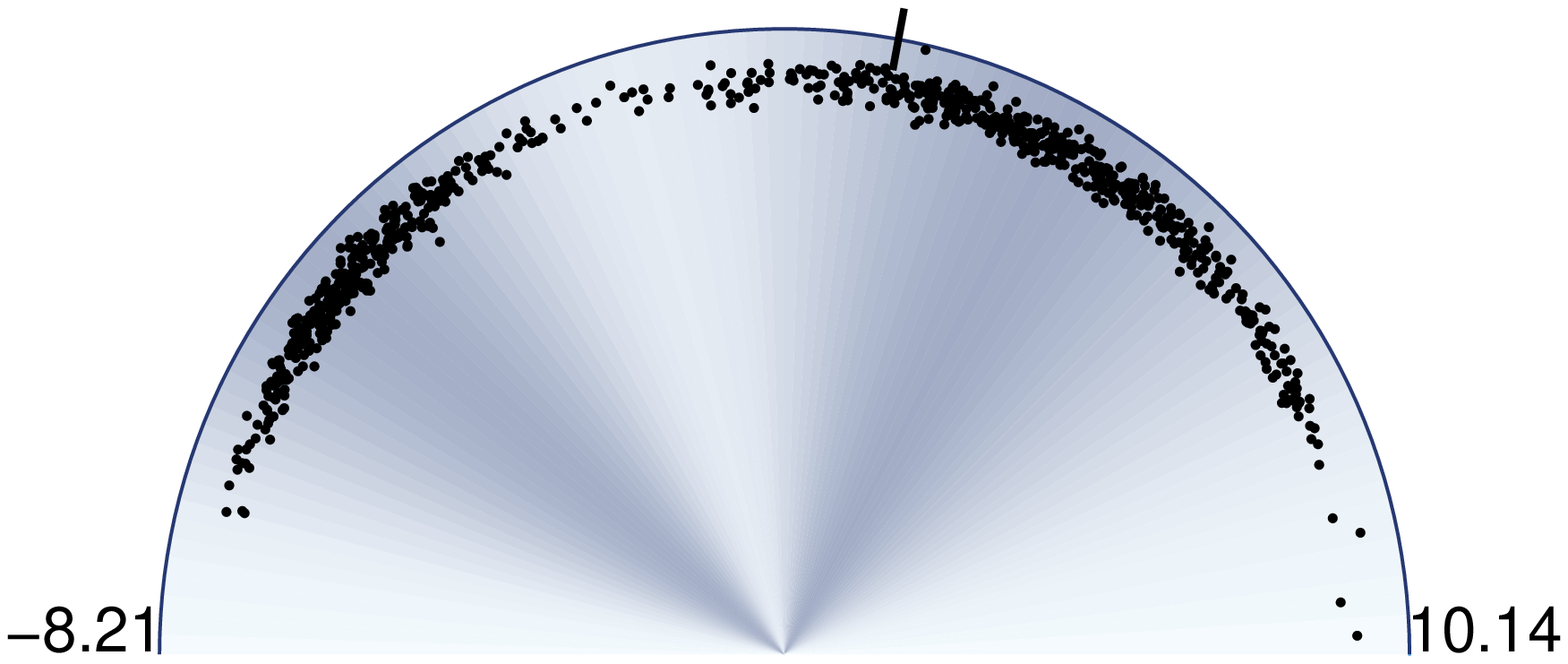}
\end{tabular}
\caption{Examples of half-disk density density strip (HDDS): discrete distribution (top-left), continuous distribution (top-right), discrete distribution with data superimposed (bottom-left) and bimodal continuous distribution (bottom-right).
Black dots represent data points.}
\label{fig:HDDS_data}
\end{figure}

Figure~\ref{fig:HDDS_data} illustrates the HDDS for different types of densities.
The red-shaded plots in the left column show the HDDSs of two samples from a negative binomial distribution. Each bin corresponds to an integer, from the smallest to the highest observations, thus allowing to easily distinguish the density associated to each value.
The bottom-left plot also reports the data by superimposing black dots to the HDDS. For visualization purposes, since repeated values from discrete distributions occur with positive probability, we have slightly perturbed the original data. As a consequence, in each bin of the strip where more than one observation falls in, the HDDS shows a small cloud of points.
The right column of Figure~\ref{fig:HDDS_data} considers two continuous distributions on the real line. The HDDS of the Gamma distribution (top) shows clear signals of skewness, having the darkest shading (i.e., probability mass) concentrated on the left side. The bottom panel shows the HDDS of a mixture of two Gaussian distributions with data points superimposed. By looking at the shading, we find evidence of bimodality with different degrees of dispersion around each mode, as is confirmed by looking at the data points.


\begin{figure}[H]  
\centering
\setlength{\tabcolsep}{5pt}
\setlength{\abovecaptionskip}{0pt}
\hspace*{-3.5ex}
\includegraphics[trim= 70mm 5mm 70mm 0mm,clip, height=3.5cm, width=4.0cm]{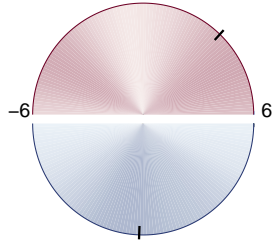}
\caption{Example of comparison between two continuous densities via HDDSs: bimodal (top, red shading) and unimodal (bottom, blue shading).}
\label{fig:HDDS}
\end{figure}

%
%
Half-disk density strips can be efficiently stacked for allowing simple pairwise comparisons of distributions. To this aim, Figure~\ref{fig:HDDS} vertically stacks two HDDSs in order to form a disk.
The bounds of the two HDDSs have been fixed to the same value for easing the comparison (although different choices can be made) and the orientation of each arc of the HDDS is set such that the lower bound corresponds to the left end and the upper bound to the right end.
In this example, the density on top is bimodal, with most of the observation lying on the rightmost part of the interval $(-6,6)$, as proved by the location of the median. Instead, the distribution on bottom has a single mode close to the midpoint of the interval.

Alternative stacking schemes can be used to visualize more than two HDDSs, however the disk-stacking as in Figure~\ref{fig:HDDS}, as opposed to standard vertical stacking, has the advantage of favoring the comparison of the length of the diameter of the HDDSs.
This is important especially in the analysis of HDDS tables, where the diameter of an HDDS is proportional to the (joint) probability of the conditioning variables, as described in Section~\ref{sec:HDDS_table}.
%

HDDSs and classical density estimation tools share two common limits.
First, they are not well suited for very small samples, since few data points do not allow for a precise nonparametric estimation of a distribution.
Second, including outliers, which by definition are located far away in the tails of the distribution, may result in hardly understandable HDDS or density plots, since these tools are not designed to detect data points in the tails.
Graphical devices which represent data points with marks are better alternatives in circumstances when one is interested in visualizing outliers.


We stress that, as for density strips, the main purpose of the half-disk density strips is to provide an intuitive tool for visualising the shape of a distribution, and not to provide the precise value of the density at every point. Therefore, the use of color shading is adequate even though perceiving fine differences in colour can be more difficult than perceiving differences in position, length, area, and volume.

\begin{rem}
Despite this paper focuses on visualizing densities of observed variables, as typical in exploratory data analysis, the HDDS can be applied to regression problems to report the distribution of an estimator (in a frequentist approach) or the posterior distribution of a parameter (in a Bayesian approach).
In this respect, it can be used as the standard density strip, as explained in \cite{bowman2019graphics}.
\end{rem}

%

\section{Half-disk density strip tables}        \label{sec:HDDS_table}
The main motivation underneath the display of numerical tables in statistics, including contingency tables, is to make comparisons.
However, as pointed out by \cite{gelman2002let}, well-designed graphs can be superior to tables for this purpose.
They suggest that data displays in statistical research should (1) identify the key comparisons of interest, (2) display these on small individual plots, and (3) establish enough control over the graphical display so that small legible plots can be juxtaposed as necessary.

In agreement with this perspective, this section shows how to use the half-disk density strip to generalise the concept of contingency table and to improve its capacity to convey information. The new graphical device, called \textit{HDDS table}, greatly boosts the capacity to make comparisons between marginal and conditional distributions of a 3-dimensional random vector.


Given a 3-dimensional random vector $X=(z,x,y)$, where $z$ is continuous, consider the problem of visualising and comparing the marginal distribution of $z$ and its conditional distributions stemming from a decomposition of the joint distribution  $p(z,x,y)$.
First, consider the case where $x,y$ are two categorical or discrete random variables.
The HDDS table can be used to represent the marginal $p(z)$ and the conditional distributions $p(z|x=x_i)$, $p(z|y=y_j)$, and $p(z|x=x_i,y=y_j)$, where $i,j$ index the values or categories of $x$ and $y$, respectively.
If instead $x$ and/or $y$ are continuous-valued, we aim at visualising the conditional distributions $p(z| x \in A_x)$, $p(z| y \in A_y)$, and $p(z| x \in A_x, y \in A_y)$, where $A_x$, $A_y$ are non-atomic subsets of the domain of $x$ and $y$, respectively.

The HDDS table is strictly related to a standard tool for exploratory data analysis of categorical variables, the contingency table.
A contingency table displays the (multivariate) joint, marginal, and conditional probability distributions of two categorical variables, using a matrix representation or a mosaic plot (see \cite{chen2007handbook}).
They are widely used before carrying statistical tests to get a visual evidence supporting the existence of an association between two variables.
%
The HDDS table extends the contingency table in several aspects.
First, a contingency table generally reports two random variables, either categorical or discrete. By contrast, an HDDS table efficiently displays information about three-dimensional (and potentially more) random vectors, and is not restricted to categorical or discrete variables.
Second, an HDDS table provides an effective tool for the visualisation of probability density functions, by means of a series of HDDSs.
This provides a more intuitive way to display the information content of the data even for a non-technical reader. By contrast, each cell of a contingency table reports the frequency of the corresponding event, thus requiring more effort to identify the key features of the distribution(s).

For example, consider a 3-variate random vector $X=(z,x,y)$, where $z$ is continuous and $x,y$ are categorical random variables which can take two possible values, $x_1$ or $x_2$ and $y_1$ or $y_2$, respectively.
Table~\ref{tab:scheme_HDDS_contingency_table} illustrates the scheme of the HDDS table for $X$ and of the contingency table for the sub-vector $(x,y)$.


\begin{table}[H]  
\centering
\hspace*{-2.5ex}
\setlength{\abovecaptionskip}{3.0pt}
\captionsetup{width=0.9\linewidth}
\subfloat[][HDDS table for $(x,y,z)$]{
\begin{adjustbox}{width=0.497\linewidth, center=0.497\linewidth}
\begin{tabular}{cc | c}
$p(z|x=x_1,y=y_1)$ & $p(z|x=x_1,y=y_2)$ & $p(z|x=x_1)$ \\
$p(z|x=x_2,y=y_1)$ & $p(z|x=x_2,y=y_2)$ & $p(z|x=x_2)$ \\
\hline
$p(z|y=y_1)$       & $p(z|y=y_2)$       & $p(z)$
\end{tabular}
\end{adjustbox}
}
\hspace*{10pt}
\subfloat[][Contingency table for $(x,y)$]{
\begin{adjustbox}{width=0.497\linewidth, center=0.497\linewidth}
\begin{tabular}{cc | c}
$p(x=x_1,y=y_1)$ & $p(x=x_1,y=y_2)$ & $p(x=x_1)$ \\
$p(x=x_2,y=y_1)$ & $p(x=x_2,y=y_2)$ & $p(x=x_2)$ \\
\hline
$p(y=y_1)$       & $p(y=y_2)$       & $1.00$
\end{tabular}
\end{adjustbox}
}
\caption{Panel (a): scheme of the HDDS table for a 3-dimensional random vector $X=(z,x,y)$, where $z$ is continuous, while $x,y$ are categorical taking two possible values.
Panel (b): scheme of the standard contingency table for a 2-dimensional random vector $Y=(x,y)$, where $x,y$ are categorical variables taking two possible values.}
\label{tab:scheme_HDDS_contingency_table}
\end{table}

Another important advantage of the use of HDDS tables over contingency tables directly refers to the way an HDDS is constructed. In fact, differently from the standard density strip (\cite{jackson2008displaying}, \cite{bowman2019graphics}), the HDDS exploits another parameter, i.e., the area of the half-disk, to visualise an additional layer of information. This becomes evident and meaningful in an HDDS table, where a set of HDDSs are contemporaneously displayed.
In this situation, the area of each HDDS is used to represent the joint and marginal probabilities of $x,y$, that is $p(x,y)$, $p(x)$, $p(y)$, by specifying a direct relationship between the area of each half-disk and the corresponding probability.
Let $I,J$ be the number of possible values of $x$ and $y$, respectively. The \textit{diameter scaling} of the HDDS in cell $(i,j)$, for $i=1,\ldots,I+1$, $j=1,\ldots,J+1$, of an HDDS table is defined as
\begin{equation}
d_{ij} = \underline{d} \cdot p_{ij}^k,
\end{equation}
where $k \in \Re_+$ is a scaling factor, $\underline{d} \in \Re_+$ is a baseline diameter and $p_{ij}$ is defined as the probability
\begin{equation*}
p_{ij} = \begin{cases}
p(x=x_i, y=y_j)  & \text{ if } i \in [1,I] \wedge j \in [1,J] \\
p(x=x_i)         & \text{ if } i \in [1,I] \wedge j = J+1 \\
p(y=y_j)         & \text{ if } i = I+1 \wedge j \in [1,J] \\
1                & \text{ if } i = I+1 \wedge j = J+1
\end{cases}
\end{equation*}
Therefore, the diameter (hence the area) of the HDDS in each cell of panel (a) in Table~\ref{tab:scheme_HDDS_contingency_table} is directly related to the probability of the conditioning variables
By applying this diameter scaling to all the HDDSs in the table, it is possible to efficiently display the information about the joint (and marginal) density of the conditioning variables in a single plot.
The rescaling is an important device for improving the readability of the HDDS table, since it allows the reader to make comparisons between its entries.



\begin{figure}[H]  
\centering
\footnotesize
\setlength{\abovecaptionskip}{4pt}
\setlength{\arraycolsep}{-1pt}
\setlength{\tabcolsep}{-4pt}
\renewcommand*{\arraystretch}{0.1}
\hspace*{-2.5ex}
\includegraphics[trim= 0mm 0mm 0mm 0mm,clip,height= 8.0cm, width= 17.0cm]{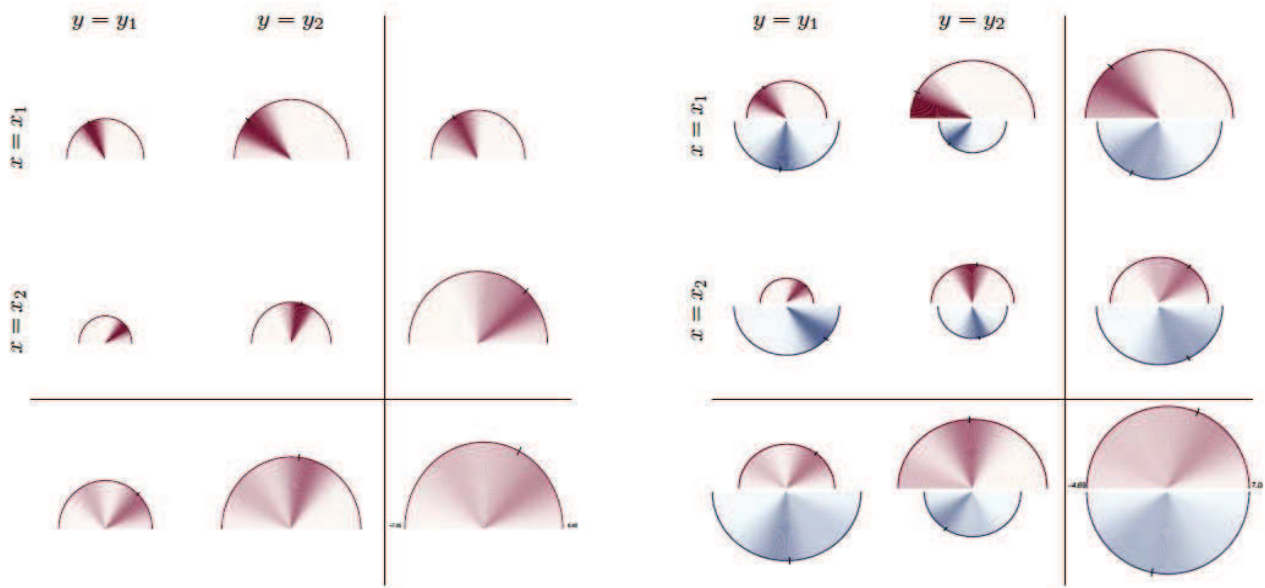}
\caption{HDDS table for a 3-dimensional vector $X=(z,x,y)$, with $z$ continuous, and $x,y$ categorical. Panel (a): single HDDS table for country 1. Panel (b): comparison of two HDDS tables, red HDDSs correspond to country 1, blue HDDSs correspond to country 2.
}
\label{fig:HDDS_table}
\end{figure}

The usefulness of the diameter scaling is shown in Panel (a) of Figure~\ref{fig:HDDS_table}, which plots an HDDS table for a 3-dimensional random vector $X=(z,x,y)$, with $z$ continuous, and $x,y$ categorical.
By looking at the bottom-right HDDS, corresponding to the marginal distribution of $z$, we find evidence of a two modes and negative skewness. A visual inspection of the last row of the table shows that the marginal distribution of $y$ is not uniform, but has more mass on $y_2$, as results from the smaller area of the bottom-left HDDS compared to the bottom-central plot.
A similar comment holds for the marginal distribution of $x$, in the rightmost column.
Moreover, by inspecting the central part of the table one finds that the smallest HDDS is in position $(2,1)$, corresponding to the distribution of $z|x=x_2,y=y_1$, and the greatest one is in $(1,2)$. This indicates that the joint distribution $p(x,y)$ has most of its mass on $(x=x_1,y=y_2)$ and just little mass on $(x=x_2,y=y_1)$.
Finally, in this example the HDDS table provides evidence about the origin of the bimodality of the marginal distribution of $z$. The conditional distributions of $z$ show that different  values of $x$ are associated to different modes of the (conditional) distribution of $z$, thus implying that, upon integrating out $x$ to obtain either $p(z|y)$ or $p(z)$, the resulting distribution is bimodal.

An additional interesting feature of HDDS tables is its ability to compare the distribution of the variables of interest from multiple datasets, e.g., over time or across geographic areas.
This is a key feature and a powerful new tool that helps scientists both in carrying exploratory data analyses and in delivering the main results to (potentially non-technical) audience.
For instance, consider the same situation analysed in the previous example and suppose data from another source, e.g. another country, is available.
By simple superposition of the country-specific HDDSs in each cell of the HDDS table, one obtains Panel (b) of Figure~\ref{fig:HDDS_table}. For each country (represented by the color), the same interpretation as in Panel (a) holds. In addition, each cell of Panel (b) provides a powerful tool for comparing distributions across countries.
Consider the first row of the HDDS table, one observes that the HDDS corresponding to $p(z|x=x_1,y=y_1)$ for country 2 is slightly bigger than that for country 1, and the converse holds for $p(z|x=x_1,y=y_2)$. This simply reflects the fact that the joint probability $p(x=x_1,y=y_1)$ for country 1 is smaller that for country 2. Clearly, the same reasoning can be applied for any cell of Panel (b).
%
Notice that the shapes of the HDDSs when conditioning on $x=x_1$ and on $x=x_2$ are different, since they put mass on different subsets of the domain. Therefore, when integrating out $x$, the resulting conditional distribution $p(z|y)$ is bimodal for both countries, and more evident for country 1 since the modes are better separated.

\begin{rem}
An HDDS table can be used also to represent multivariate distributions with dimension $n > 3$, similarly to standard contingency tables.
One way to do this is by substituting the marginal distribution of one conditioning variable with the joint distribution of two or more. For example, in a 4-dimensional case one may substitute $y$ for the vector $(y,w)$. As a consequence, the interpretation of the central part and of the bottom row of the HDDS table in panel (a) of Table~\ref{tab:scheme_HDDS_contingency_table} will change accordingly.
\end{rem}

\section{Applications}     \label{sec:illustrations}

This section presents two applications of the HDDS table on real datasets.
The purpose is to shed light on the insights and benefits that a researcher can get from the use of HDDS tables. Even though the HDDS table does not provide formal statistical evidence, being a pure visualization tool, it nonetheless helps to obtain meaningful information about key characteristics of the data, that are the starting point for subsequent analyses.

We stress that the findings highlighted in this section are just the outcome of a visual inspection and do not represent formal statistical analyses. Other statistical tools, such as formal testing procedures, are required to validate the insights provided by the visualization of the data with HDDSs.

\subsection{Case study I: Italian income}

Consider the problem of visualizing and comparing the income distribution in a population, discriminating by gender and employment type.
 This is a well-discussed issue in economics (e.g., see \cite{bertrand2015gender}).

We consider data for Italy, from the Survey of Households, Income and Wealth (SHIW) collected in 2016. The distribution of the net disposable income, $z$, is reported for male and female workers, $x$, distinguished into four employment categories, $y$: blue-collar, office workers, cadre/managers and self-employed.
See Appendix \ref{sec:apdx_data} for more details.

\begin{figure}[H]
\centering
\footnotesize
\hspace*{-6.5ex}
\resizebox{0.9\textwidth}{!}{
\begin{tabular}{c c c c c | c}
 & blue-collar & office worker & cadre/manager & self-employed & \\
\begin{rotate}{90} \hspace*{24pt} male \end{rotate} &
\includegraphics[trim=5mm 5mm 0mm 4mm,clip,height= 3.0cm, width=3.3cm]{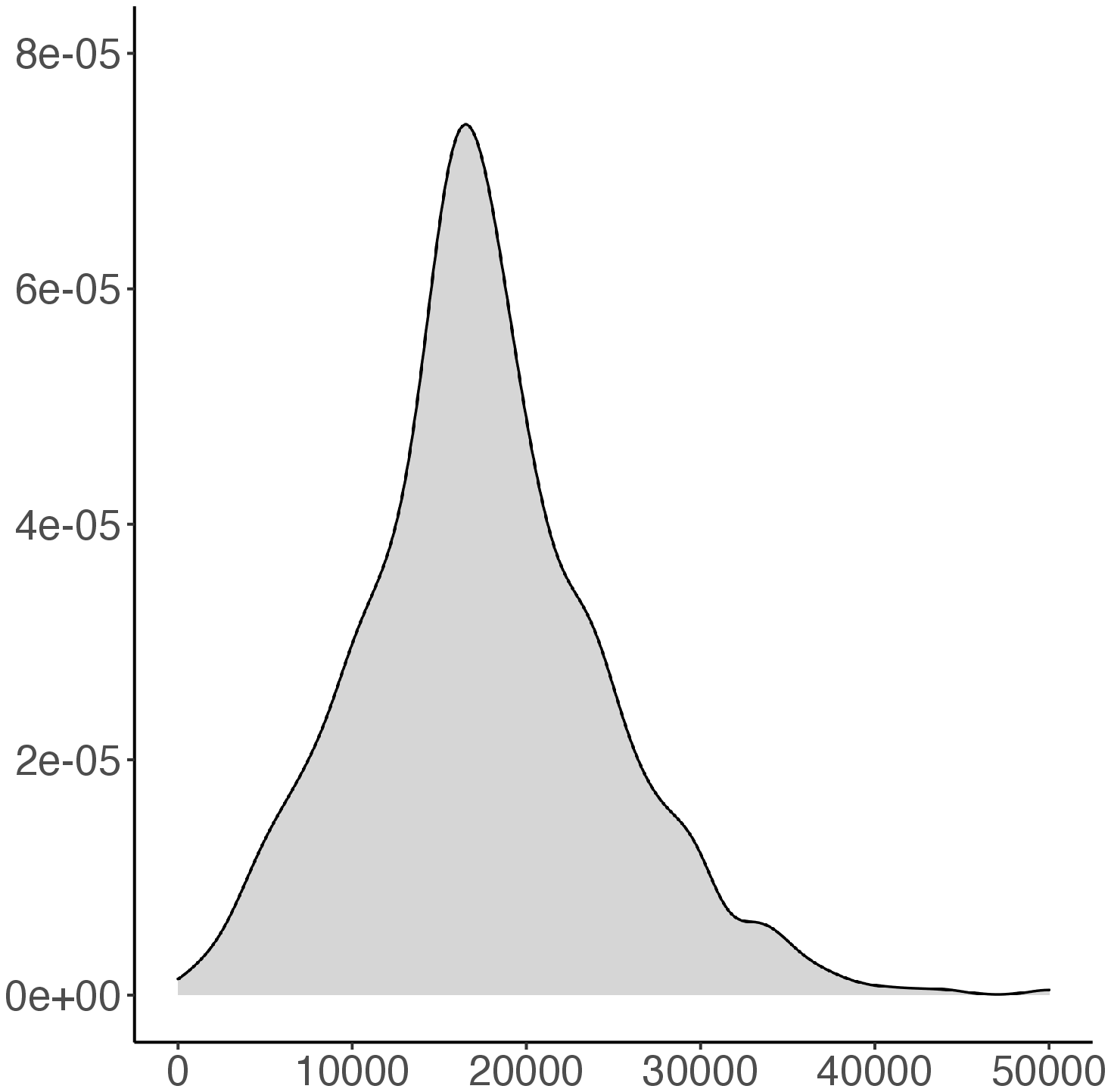} &
\includegraphics[trim=5mm 5mm 0mm 4mm,clip,height= 3.0cm, width=3.3cm]{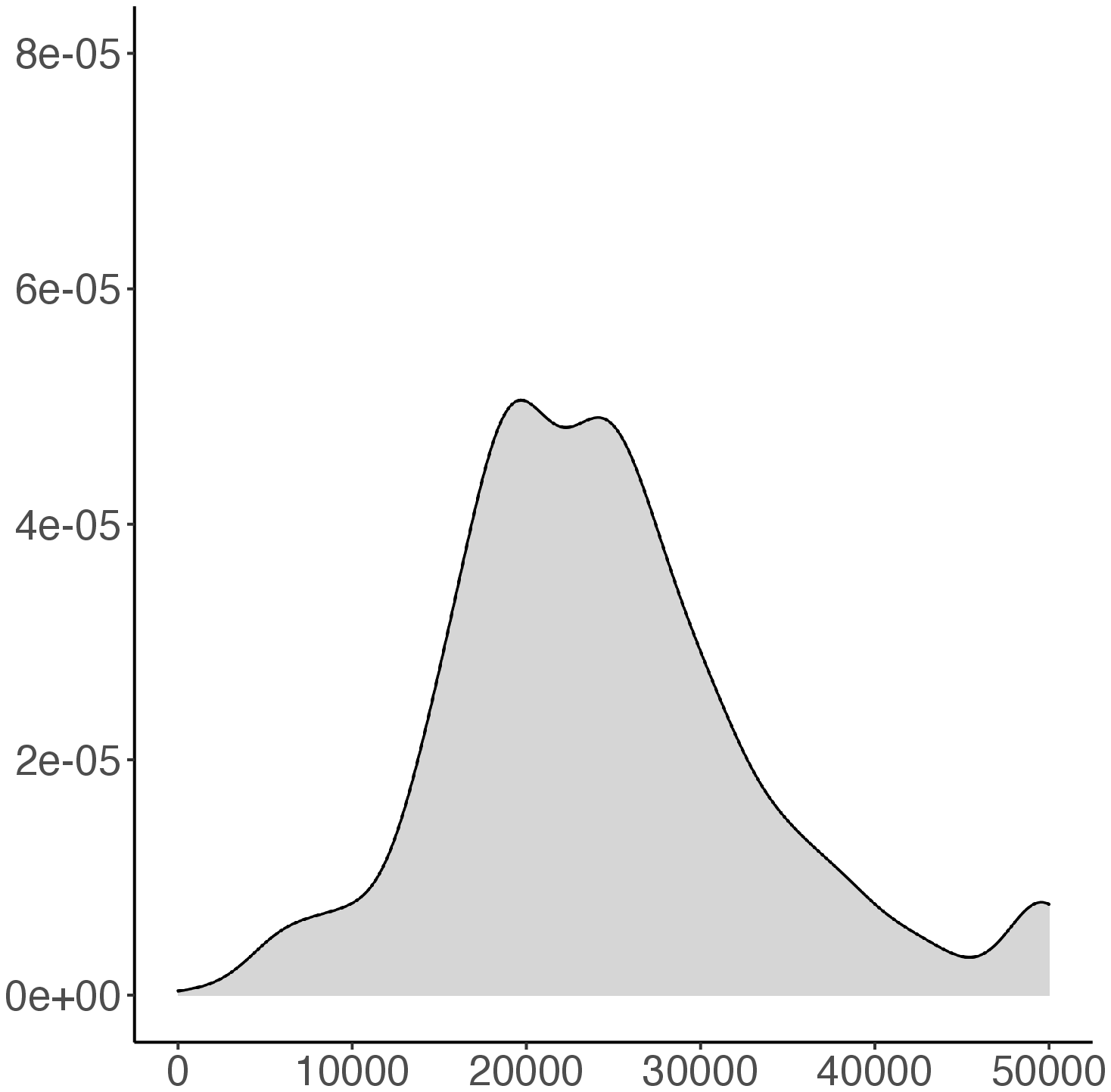} &
\includegraphics[trim=5mm 5mm 0mm 4mm,clip,height= 3.0cm, width=3.3cm]{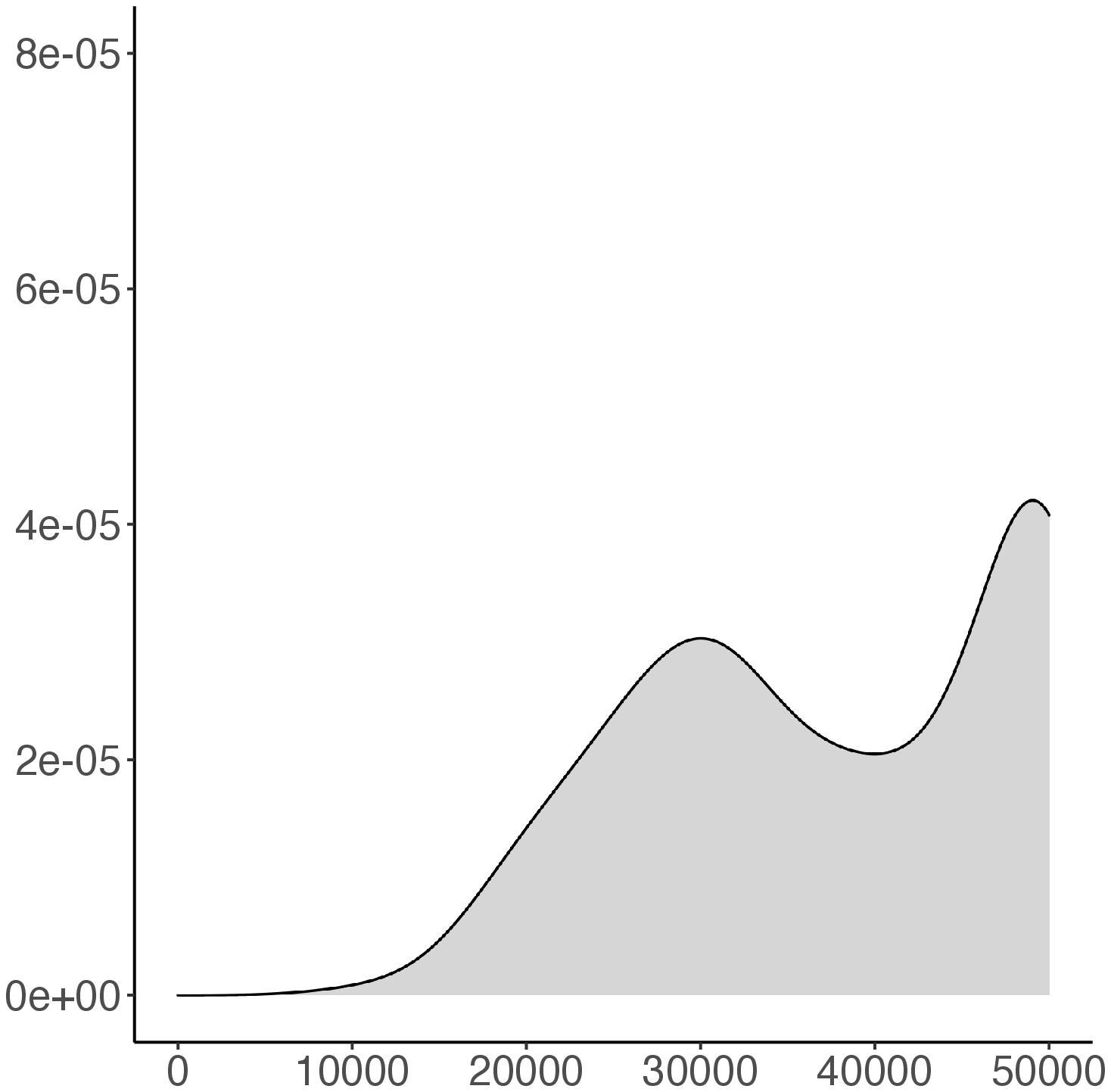} &
\includegraphics[trim=5mm 5mm 0mm 4mm,clip,height= 3.0cm, width=3.3cm]{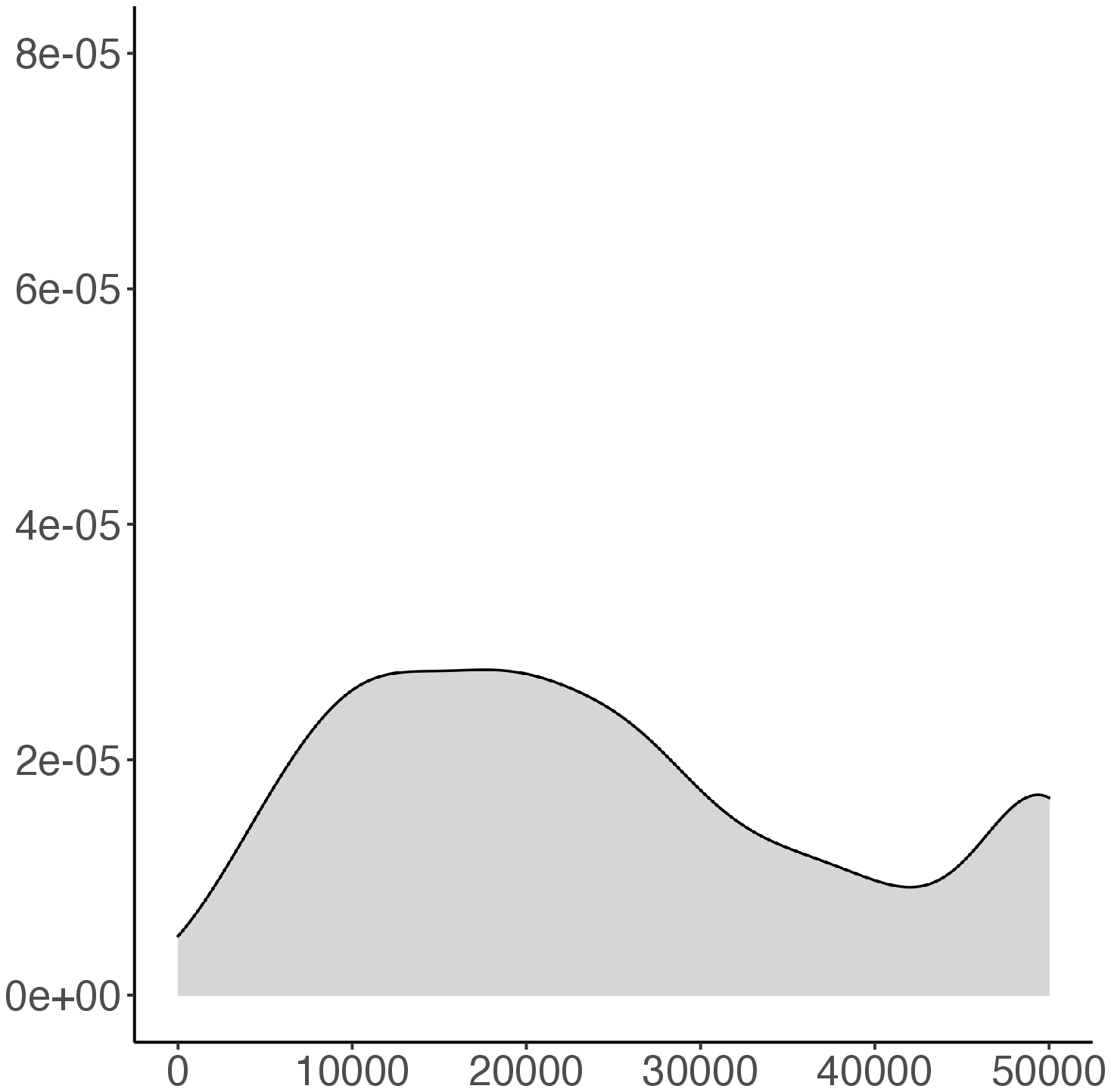} &
\includegraphics[trim=5mm 5mm 0mm 4mm,clip,height= 3.0cm, width=3.3cm]{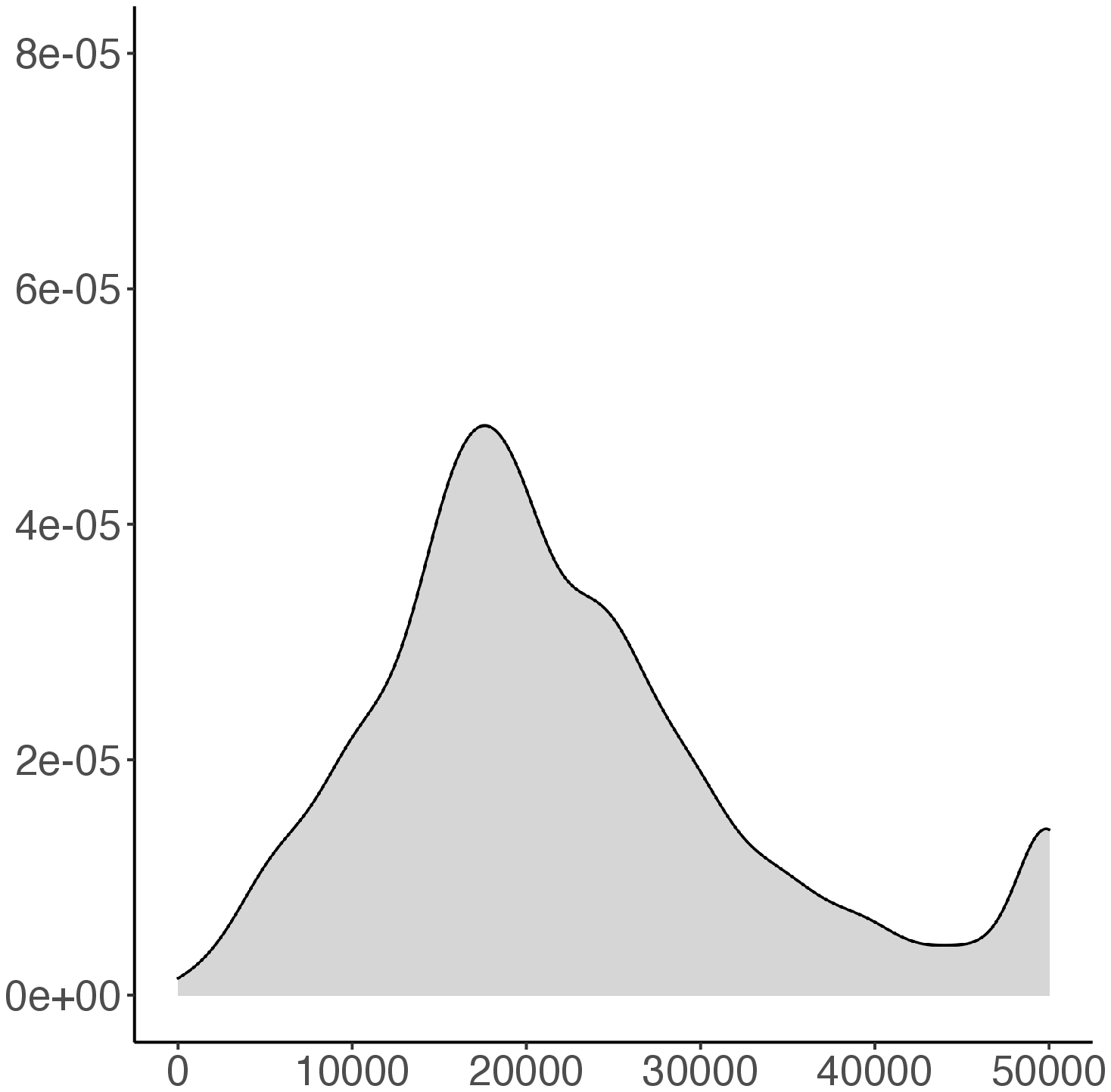} \\[6pt]
\begin{rotate}{90} \hspace*{20pt} female \end{rotate} &
\includegraphics[trim=5mm 5mm 0mm 4mm,clip,height= 3.0cm, width=3.3cm]{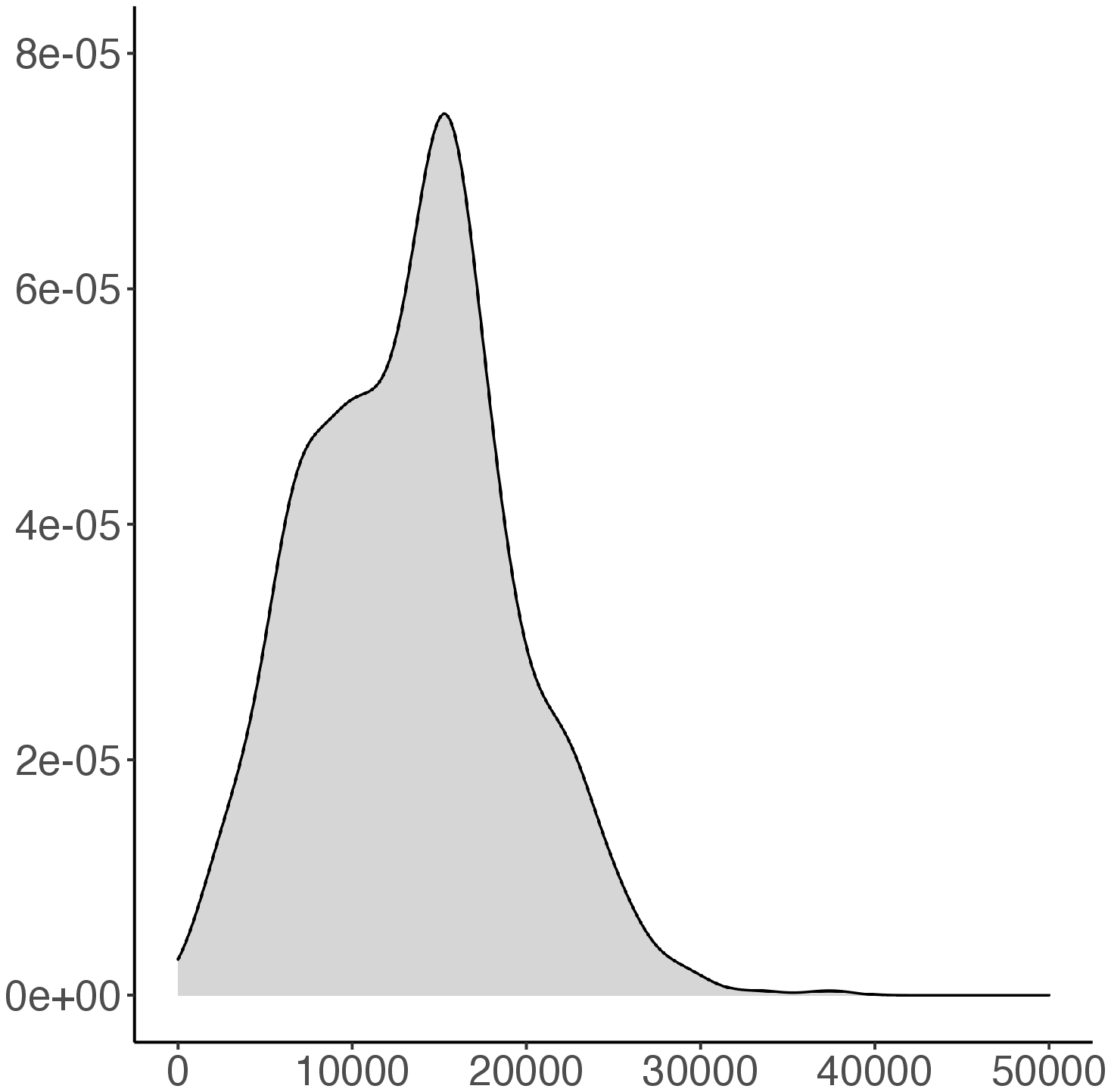} &
\includegraphics[trim=5mm 5mm 0mm 4mm,clip,height= 3.0cm, width=3.3cm]{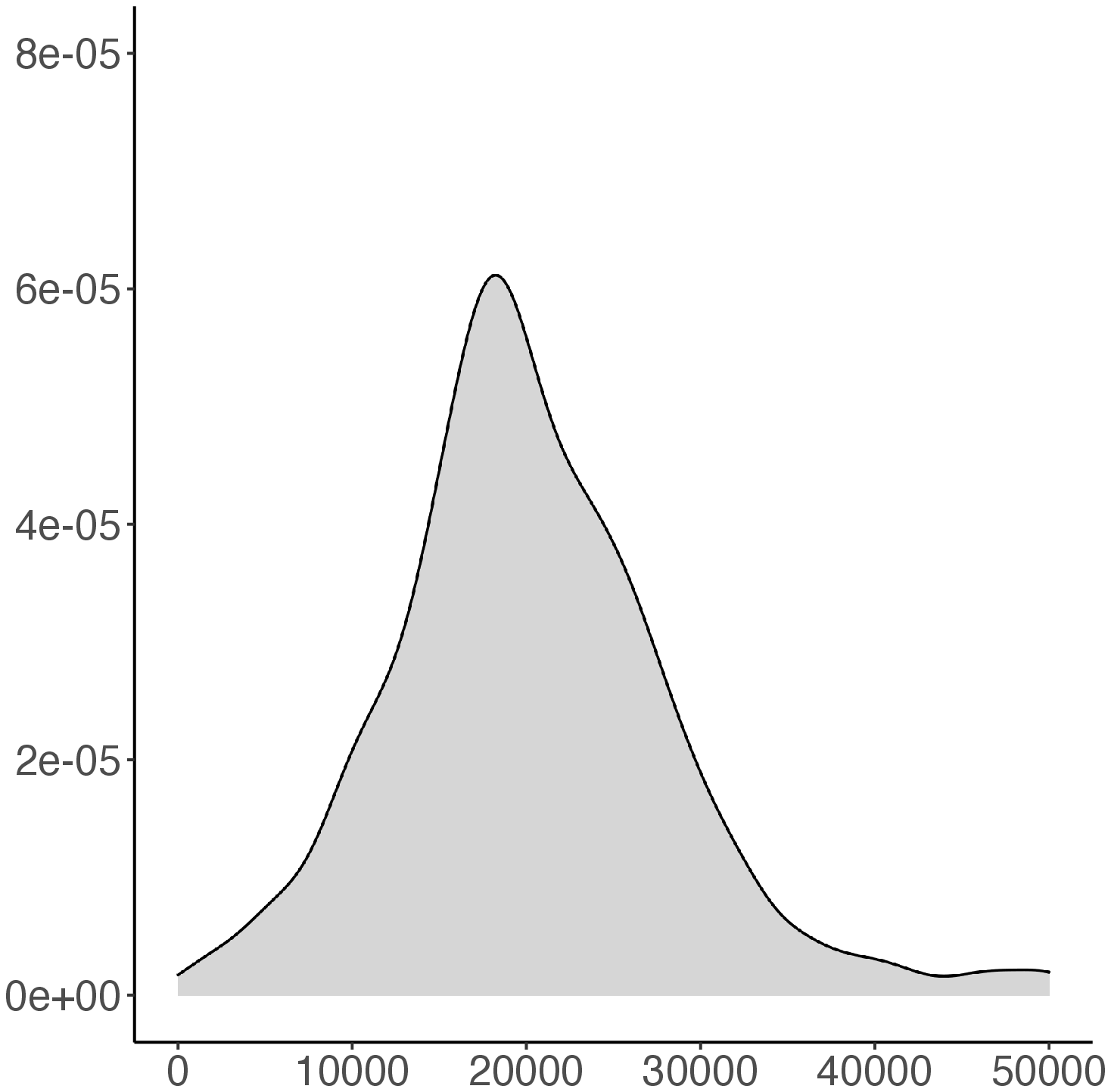} &
\includegraphics[trim=5mm 5mm 0mm 4mm,clip,height= 3.0cm, width=3.3cm]{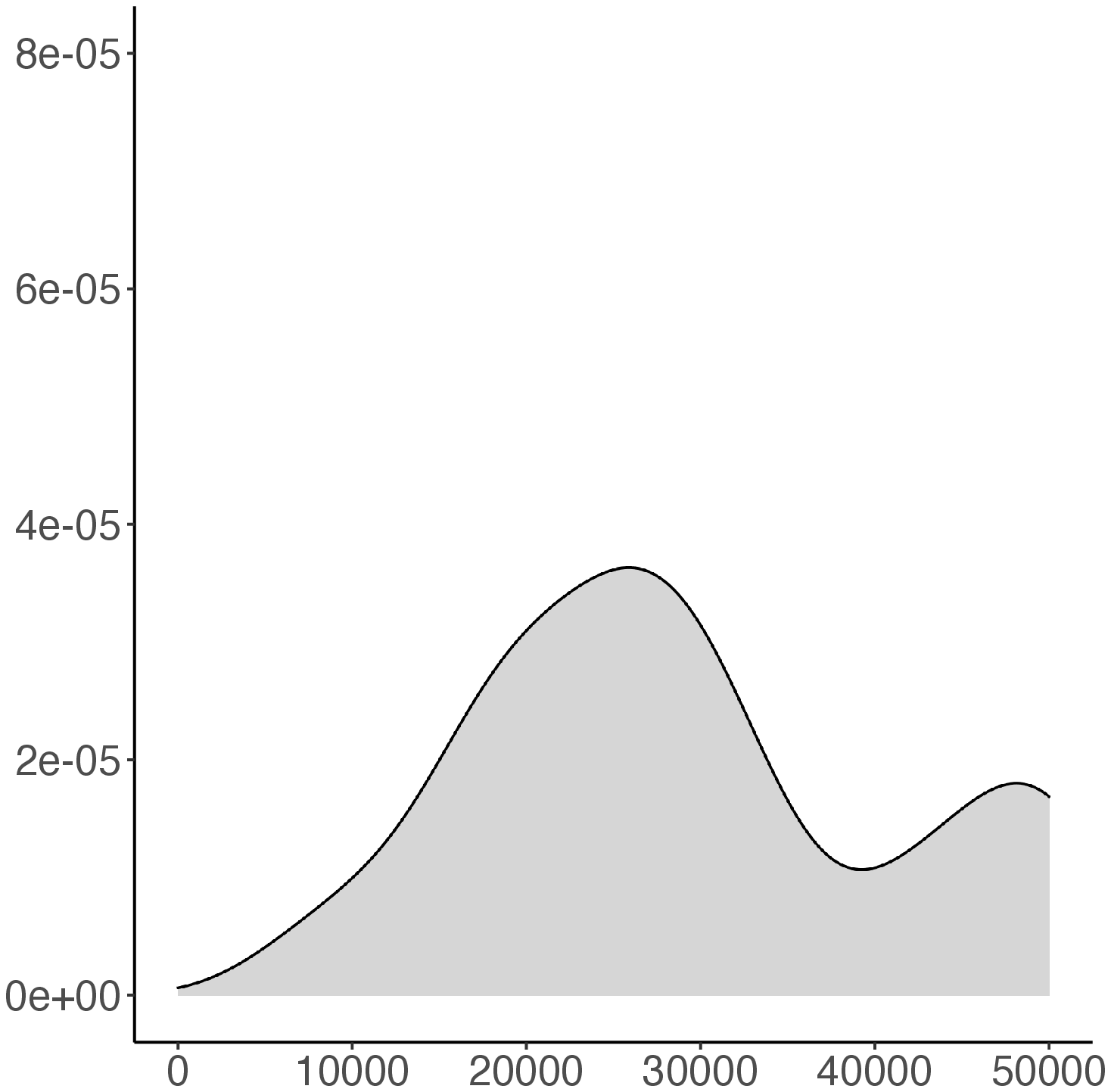} &
\includegraphics[trim=5mm 5mm 0mm 4mm,clip,height= 3.0cm, width=3.3cm]{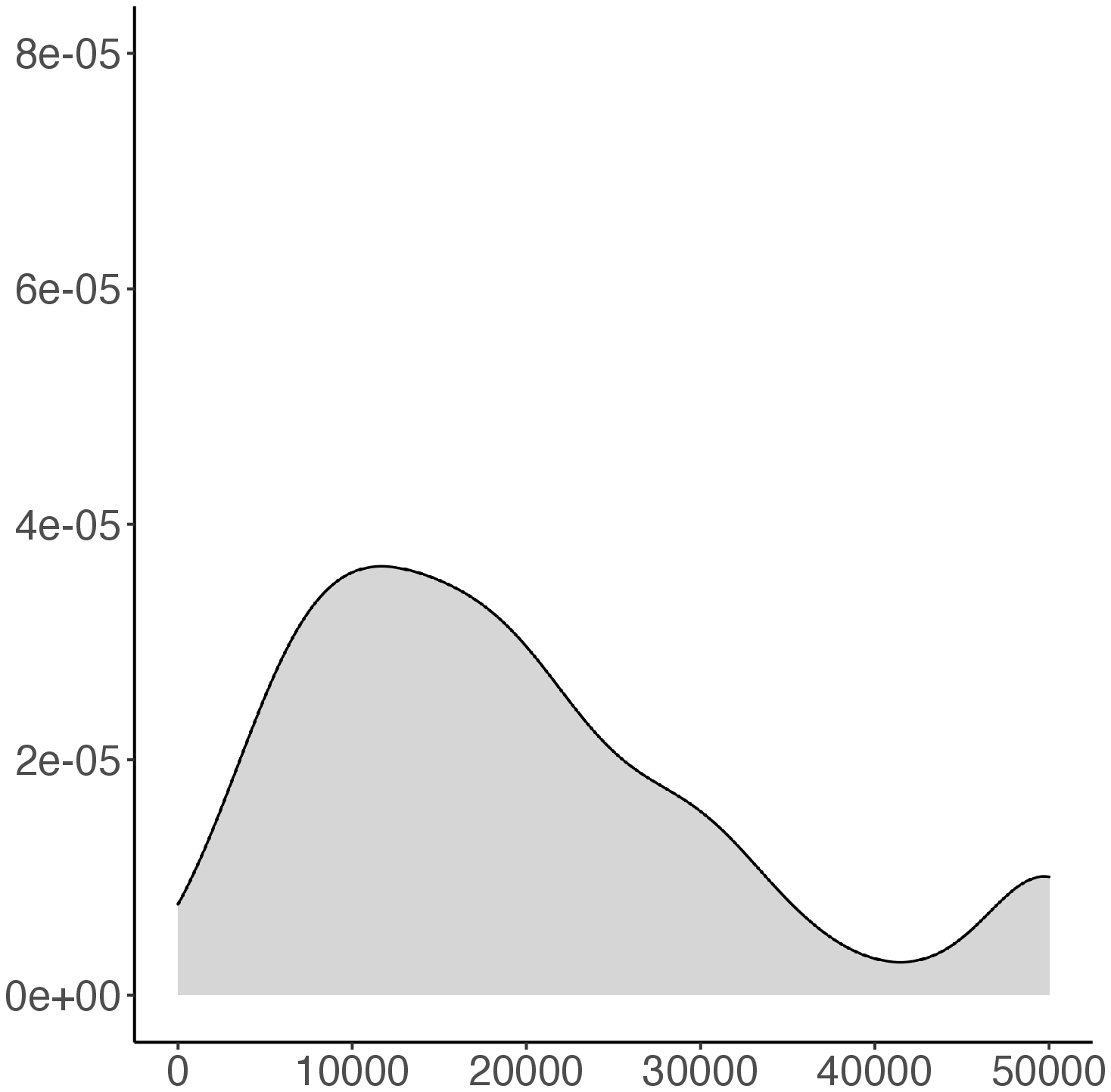} &
\includegraphics[trim=5mm 5mm 0mm 4mm,clip,height= 3.0cm, width=3.3cm]{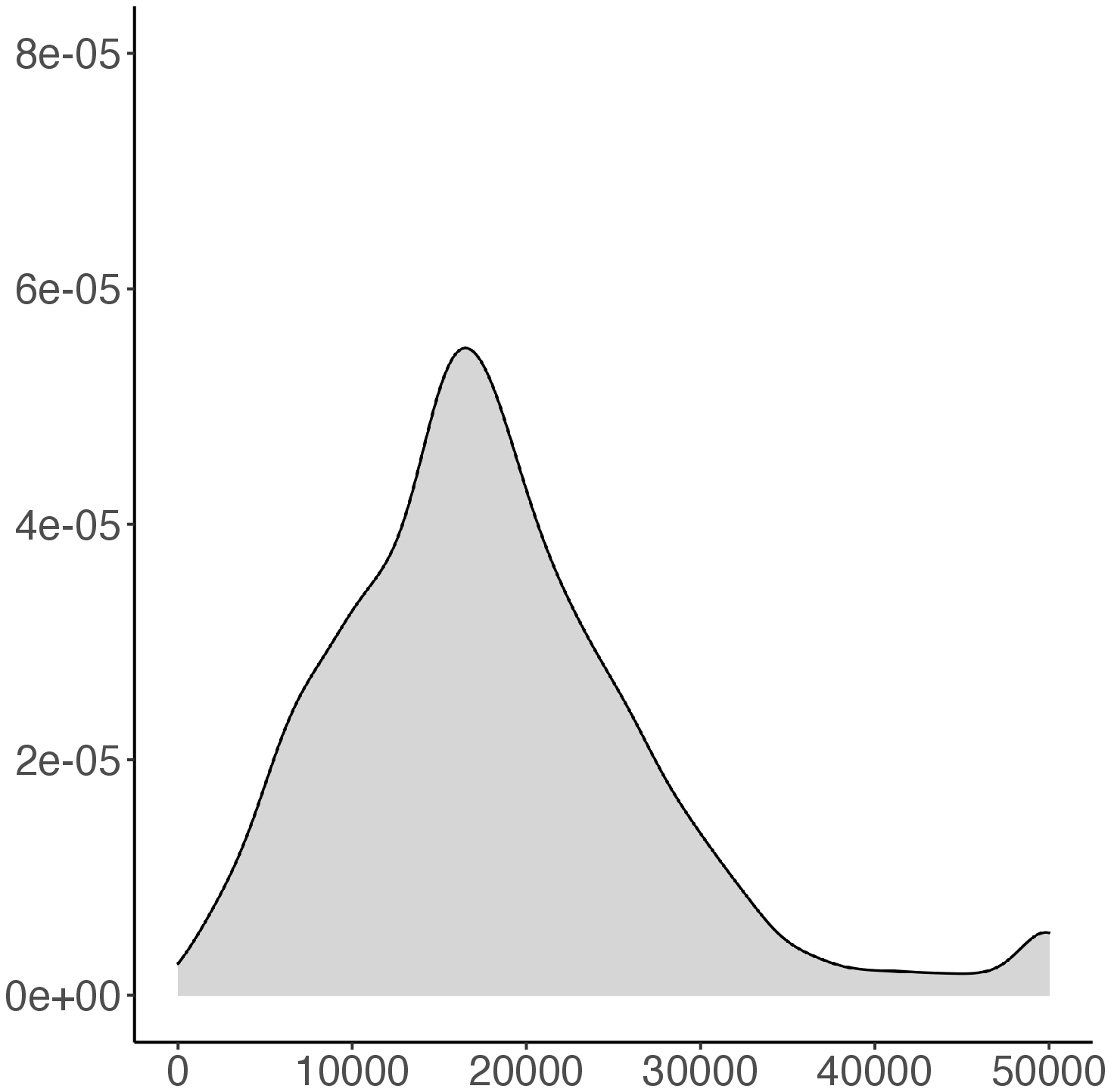} \\[6pt] \hline
 & & & & \\
 & 
\includegraphics[trim=5mm 5mm 0mm 4mm,clip,height= 3.0cm, width=3.3cm]{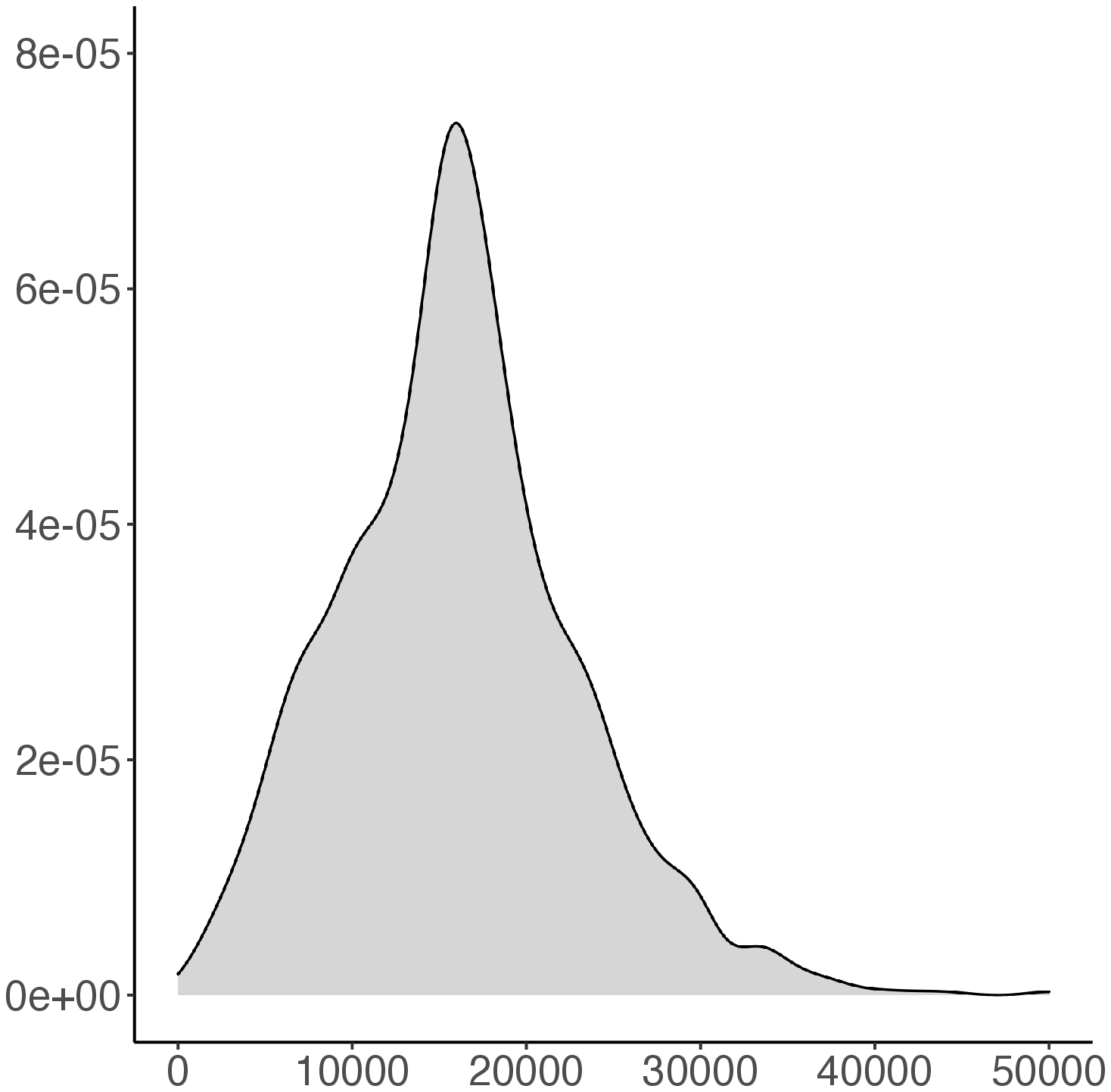} &
\includegraphics[trim=5mm 5mm 0mm 4mm,clip,height= 3.0cm, width=3.3cm]{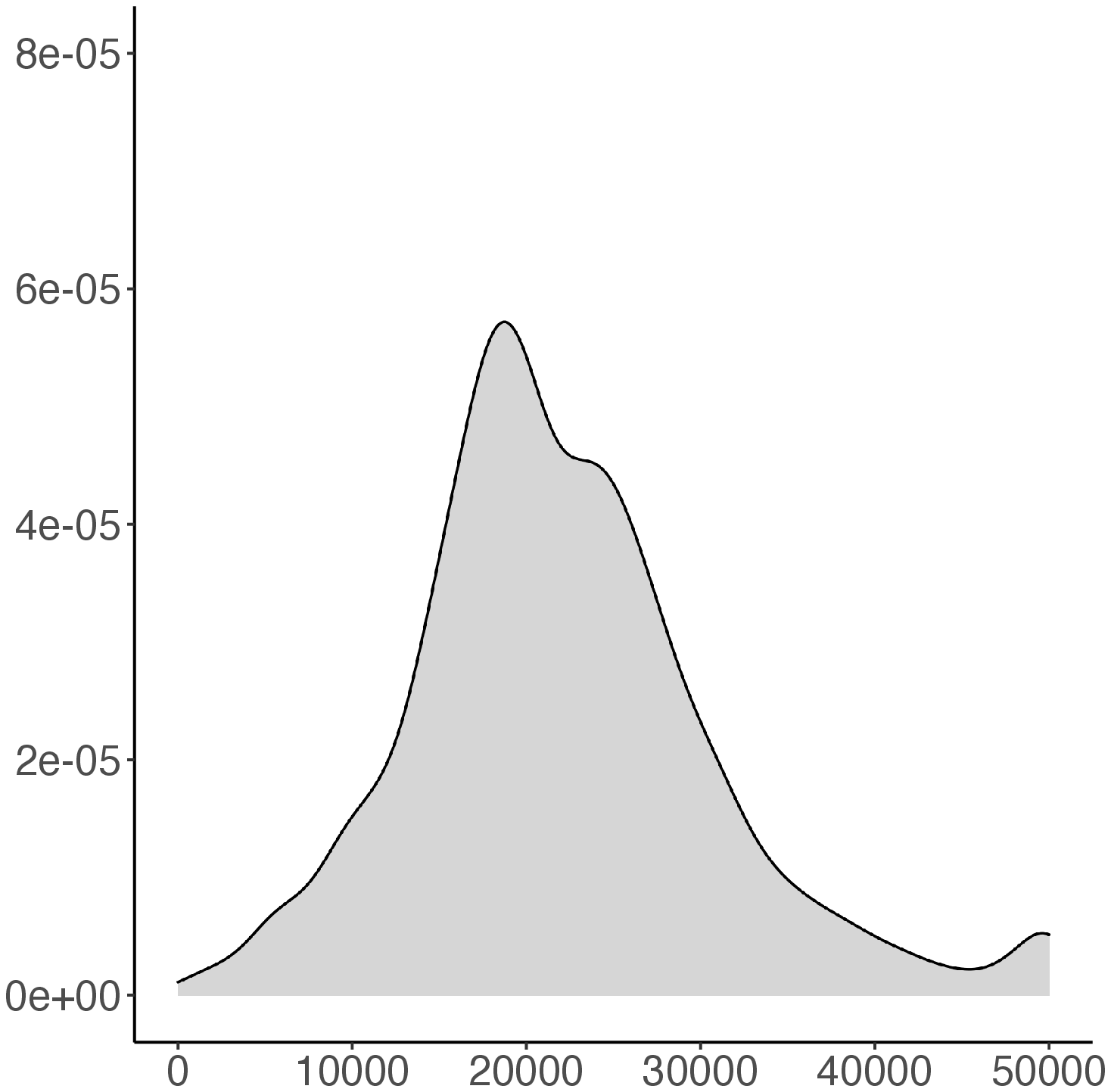} &
\includegraphics[trim=5mm 5mm 0mm 4mm,clip,height= 3.0cm, width=3.3cm]{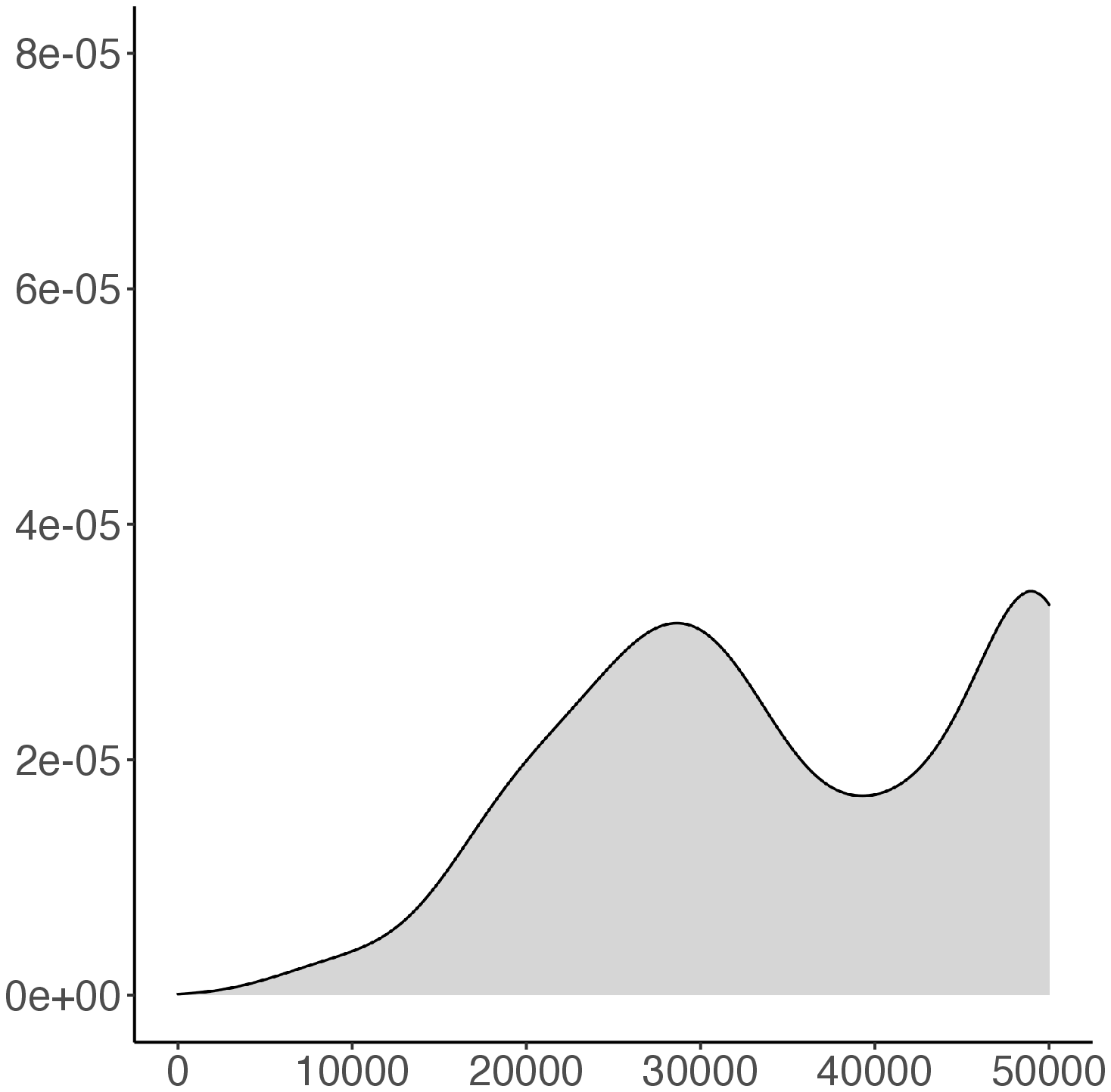} &
\includegraphics[trim=5mm 5mm 0mm 4mm,clip,height= 3.0cm, width=3.3cm]{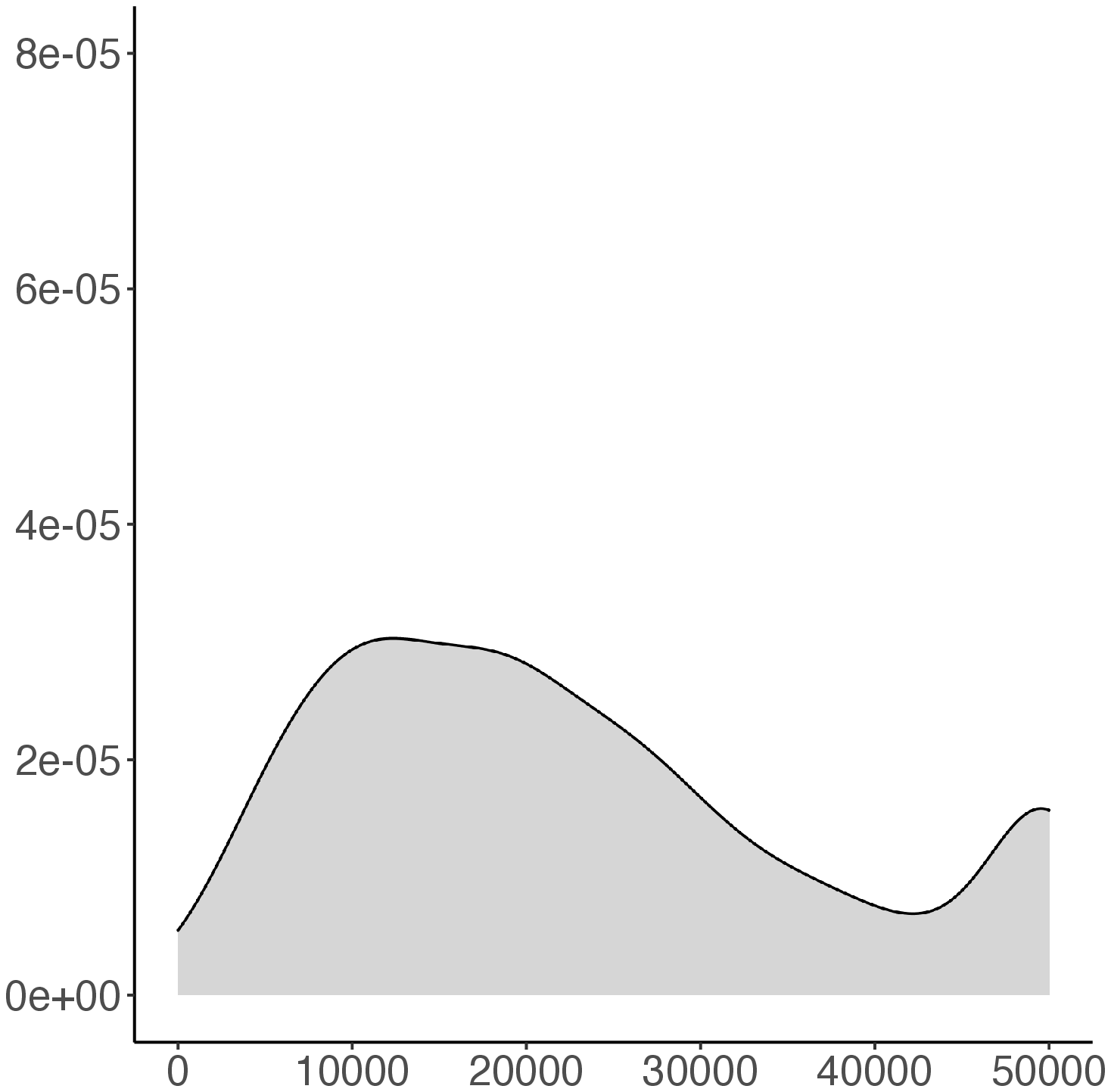} &
\includegraphics[trim=5mm 5mm 0mm 4mm,clip,height= 3.0cm, width=3.3cm]{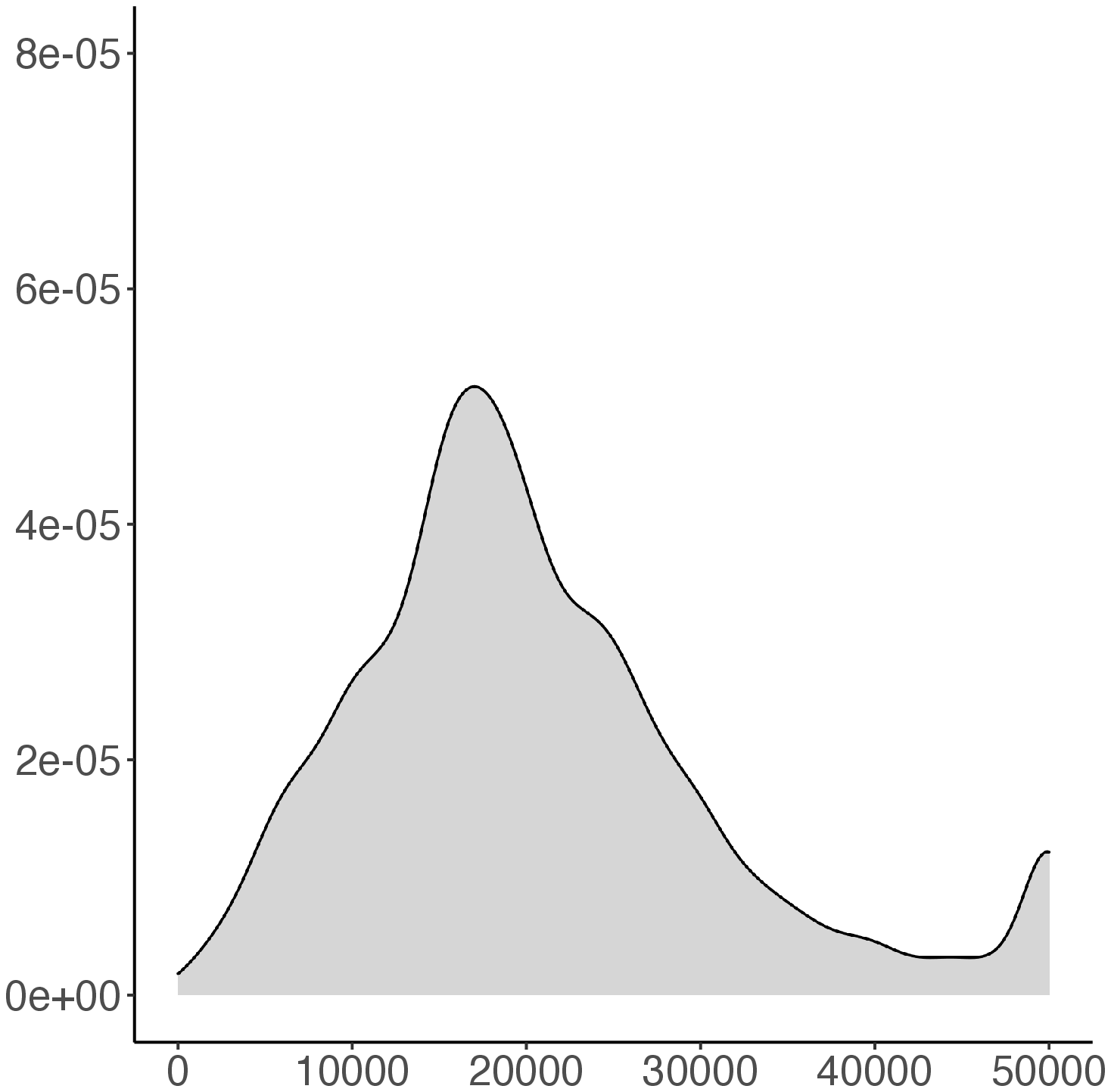}
\end{tabular}
}

\vspace*{4ex}
{\footnotesize
\renewcommand*{\arraystretch}{1.1}
\begin{tabular}{ccccc|c}
 & {\scriptsize blue-collar} & {\scriptsize office worker} & {\scriptsize cadre/manager} & {\scriptsize self-employed} & \\
\begin{rotate}{90} {\scriptsize \hspace*{-4pt} male} \end{rotate} &
0.252 & 0.151 & 0.040 & 0.135 & 0.578 \\[6pt]
\begin{rotate}{90} {\scriptsize \hspace*{-10pt} female} \end{rotate} &
0.142 & 0.201 & 0.020 & 0.060 & 0.422 \\[6pt]
\hline
 & 0.393 & 0.352 & 0.060 & 0.195 & 1.000
\end{tabular}
}
\caption{Top: table of kernel density plots for income distribution in Italy (SHIW 2016), by gender (rows) and employment status (columns).
Bottom: contingency table for the distribution of gender and employment status.}
\label{fig:SHIW_IT_income_kernel_density}
\end{figure}

Standard statistical tools for analyzing this dataset are kernel density estimators and contingency tables, which are reported in Figure~\ref{fig:SHIW_IT_income_kernel_density}.
A fundamental issue that immediately emerges concerns the bottom row and the right column, where the conditional densities $p(z|y)$ and $p(z|x)$ are shown, respectively.
These conditional distributions are obtained by integrating out one of the conditioning variables, but, by construction, the set of kernel density estimates does not provide any information about the probability distribution of the conditioning variables $x,y$.
Consider for example the distribution of income for males (first row, last column of \ref{fig:SHIW_IT_income_kernel_density}), $p(z|x)$. One might find it hard to interpret the small weight of the right tail, compared to the distribution of income for male managers (first row, third column).
The reason is that this class of workers is sensibly smaller than the others, as shown in the contingency table in the bottom panel of Figure~\ref{fig:SHIW_IT_income_kernel_density}, thus its contribution to $p(z|x)$ is minimal.
Neglecting the information about the joint and marginal probabilities of the conditioning variables $x,y$ may potentially lead to misinterpretations of the kernel density estimates, especially for a non-technical reader.


\begin{figure}[H]  
\centering
\footnotesize
\setlength{\tabcolsep}{-5pt}
\renewcommand*{\arraystretch}{0.85}
\includegraphics[trim= 0mm 0mm 0mm 0mm,clip,scale=0.45]{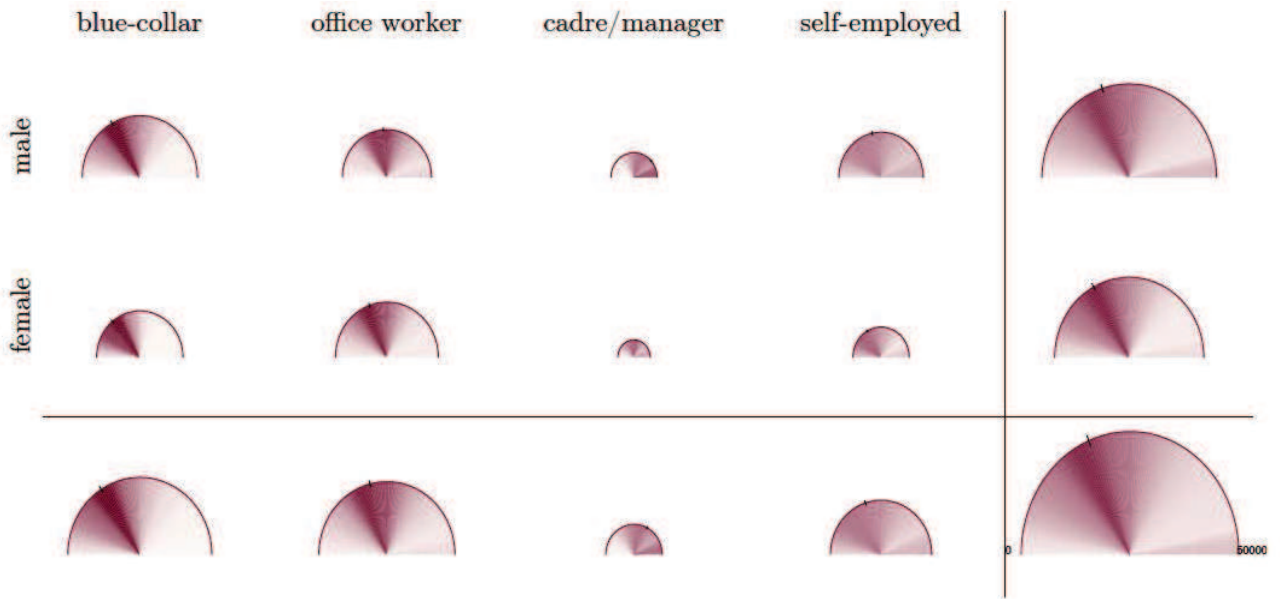}
\caption{HDDS table for the income distribution in Italy (SHIW 2016), by gender (rows) and employment status (columns).}
\label{fig:SHIW_IT_income_HDDS}
\end{figure}

On the other hand, the HDDS table in Figure~\ref{fig:SHIW_IT_income_HDDS} reports this information in an intuitive manner, by scaling the area of the HDDSs according to the joint or marginal probability of the corresponding conditioning variables.
For example, the HDDS of the income distribution for male managers is remarkably smaller than that of blue-collars, thus visually representing in a direct way the content of  the contingency table in Figure~\ref{fig:SHIW_IT_income_kernel_density}. This provides an intuitive rationale for the HDDS of the distribution of income of males in the top-right corner.
The HDDS table in Figure~\ref{fig:SHIW_IT_income_HDDS} also shows interesting results by comparing the income distribution for male and female workers. First, the area of the HDDSs indicates that females in the sample are most often office workers and the relative frequency of female blue-collar, manager and self-employed workers is lower than the corresponding shares for males. Moreover, most of the males are blue-collar workers.
Overall, this suggests an heterogeneous distribution of employment status across gender.
Second, the income distribution for male has fatter right tail and higher median as compared to females (see the rightmost column of Figure~\ref{fig:SHIW_IT_income_HDDS}), providing evidence of inequality in income distribution for the two genders.
Third, in terms of employment class, as expected, the income distribution of managers has more weight on the right tail, as opposed to blue-collar workers, whereas the income of self-employed people is more uniformly distributed (see the bottom row of Figure~\ref{fig:SHIW_IT_income_HDDS}).
Summarizing, the HDDS table is thus more informative than existing statistical tools usually employed in exploratory data analysis.

We remark that the use classical methods to visualize the information conveyed by the HDDS table in Figure~\ref{fig:SHIW_IT_income_HDDS} requires a higher number of plots and a more complex interpretation. In fact, the kernel density plots (or others) of each conditional density $p(z|x_i,y_j)$, $p(z|x_i)$, $p(z|y_j)$ and the marginal density $p(z)$, do not provide any information about the joint distribution of the conditioning variables, $(x,y)$, thus requiring additional tools, such as a contingency table, to represent it.
Moreover, the interpretation of a series of density plots for the marginal and conditional densities of $z$ and a contingency table for the density of $(x,y)$ is far from straightforward. As shown in Figure~\ref{fig:SHIW_IT_income_HDDS}, an effective visual comparison of two densities requires to weight each density plot by the relative frequency in the corresponding position of the contingency table.
For example, to correctly compare $p(z| x_i,y_j)$ and $p(z| x_k,y_j)$ it would be necessary to weight the two density plots by $p(x_i,y_j)$ and $p(x_k,y_j)$, respectively.


\begin{figure}[H]  
\centering
\footnotesize
\hspace*{-3.5ex}
\setlength{\tabcolsep}{-8pt}
\renewcommand*{\arraystretch}{0.25}
\includegraphics[trim= 0mm 0mm 0mm 0mm,clip,scale=0.45]{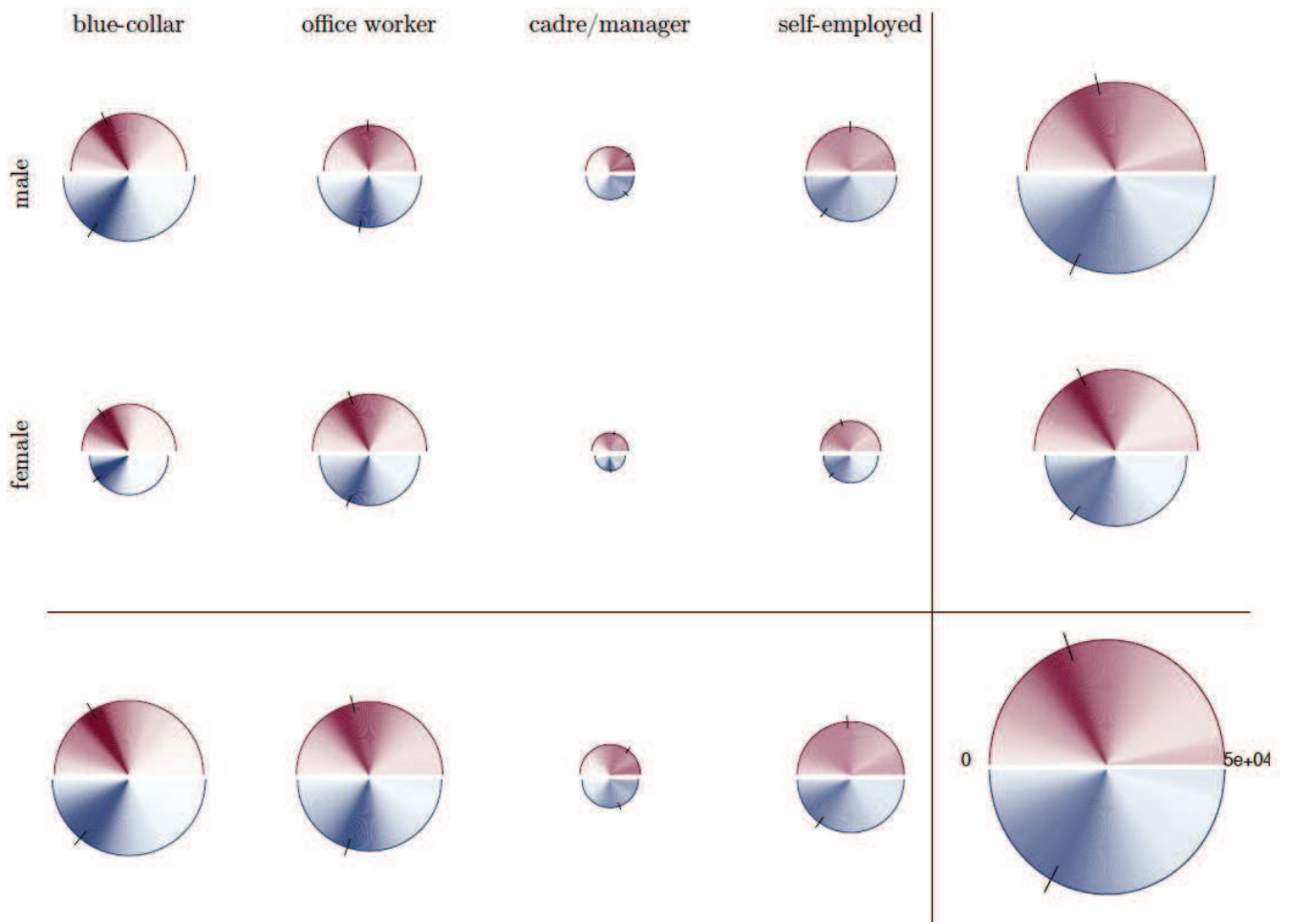}
\caption{HDDS table for the income distribution in Italy (SHIW 2016), by gender (rows) and employment status (columns), for northern regions (top, red shading) and southern regions including islands (bottom, blue shading).}
\label{fig:SHIW_IT_income_north_south}
\end{figure}

The SHIW dataset also contains geographical information about the individuals, such as the region of residence.
To deepen further the exploratory analysis of the data, we apply the HDDS table to perform a comparison of the income distribution between the Italian Northern and Southern regions in Figure~\ref{fig:SHIW_IT_income_north_south}, using red and blue shading, respectively.
We find several interesting results.
To begin with, the marginal distribution of income (bottom-right HDDS) shows that the median income is higher for Northern regions, which also are more likely to have very high incomes as implied by the darker red shade on the right tail.
When studying the composition of employment, the equal size of the areas of the HDDSs in the bottom row of the table provides no evidence of difference between the two macro-regions.
Instead, the analysis of $p(z|x)$ in the last column of  Figure~\ref{fig:SHIW_IT_income_north_south} illustrates that in the North there is a more uniform distribution of the number of male and female workers, see the almost equal area of the red-shaded HDDSs.
Another interesting feature emerges from the inspection of the inner part of the HDDS table.
The income distribution of self-employed workers geographically differs: in the North, it seems rather uniform for both males and females, while in the South the distribution is more concentrated on lower income levels, especially for female workers.
A similar differences of income distribution between North and South and between genders is found for cadres/managers.


Overall, this case study highlights the potential of the HDDS and HDDS table in conveying information and facilitating the comparison of probability distributions. It is found to be an effective tool for uncovering features of the data that may lead to further, more in-depth statistical investigation.
Moreover, HDDS tables have been proven to generalize and improve standard statistical tools, such as kernel density plots and contingency tables, while also allowing the non-technical public to easily interpret the plot.

\subsection{Case study II: life satisfaction}

In this case study, we analyse data from the Survey of Health, Ageing and Retirement in Europe\footnote{Source: \url{https://www.share-datadocutool.org/study_units/view/11}} (SHARE), wave 7, which contains socio-economic and medicare information for European individuals aged 50 years and over.

We consider the problem of visualizing self-reported life satisfaction as a function of the individual's current income class and the economic status of its family of origin.
Literature about the relation between income and life satisfaction (\cite{boyce2010money}) has evidenced the existence of a positive relationship between current income rank and life satisfaction.
We aim at exploring this relationship by considering an additional variable in our analysis, the ex-post perceived economic status of the native family during the subject' childhood (i.e., between 0-16 years).
This categorical variable proxies the wealth of an individual before the start of his working career.

\begin{figure}[H]  
\centering
\footnotesize
\setlength{\tabcolsep}{-5pt}
\renewcommand*{\arraystretch}{0.85}
\includegraphics[trim= 0mm 0mm 0mm 0mm,clip,scale=0.45]{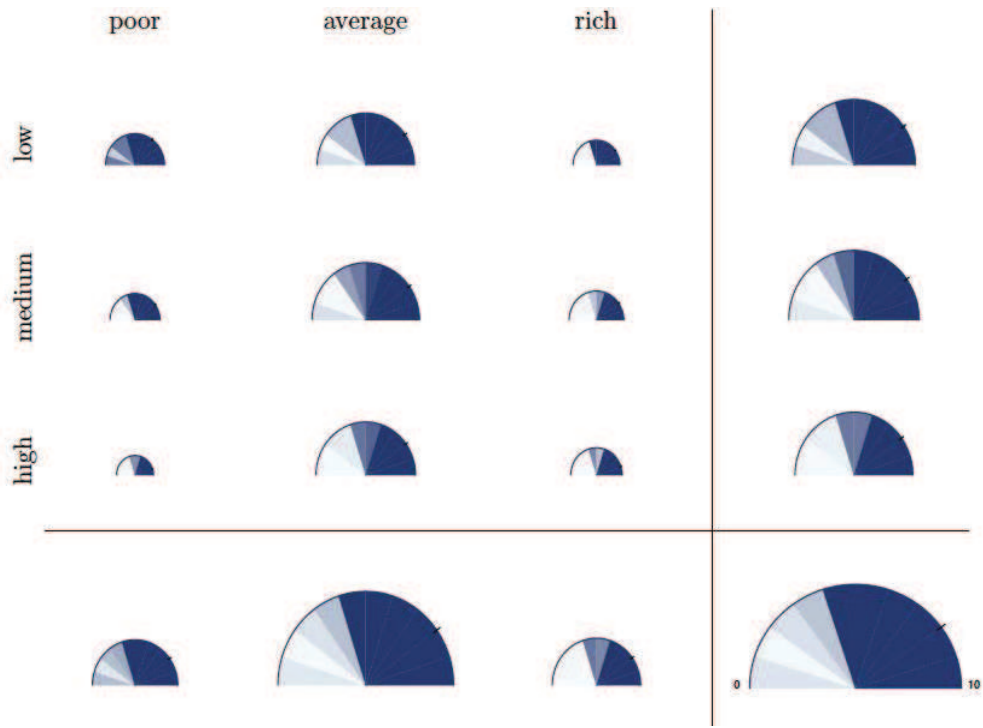}
\caption{HDDS table for life satisfaction in EU (SHARE w.7), by current monthly income (rows) and family status in 0-16 (columns).}
\label{fig:share_HDDS}
\end{figure}

In order to build an HDDS table, we first partitioned the positive real-valued current income, $x$, into three intervals defined by equally-spaced percentiles, which denote low, medium, and high income classes, respectively. Figure~\ref{fig:share_HDDS} reports the results.


Overall, we find evidence of a positive association between the level of current income and life satisfaction, as we can see from the last column of the table. In fact, for higher classes of income, the median value of life satisfaction is higher, and the whole distribution has more mass on the right tail.
Interestingly, a similar result is found when conditioning only on the family status (see the last row of Figure~\ref{fig:share_HDDS}): people coming from rich families have higher life satisfaction on average.
Moreover, by comparing the last row and last column of the HDDS table, we notice that while the majority of individuals in the sample was born in a middle-class family (see the area of the HDDS in the center of the last row), the distribution across current income levels seems to be more uniform, as stems from the similar area of the HDDS in the last column.

The HDDS table provides some interesting insights as a byproduct.
Considering the economic status of the family between 0-16 years as a proxy of parents' wealth and the current income as a proxy for an individual's wealth, then the joint distribution of these variables provides an approximate measure for social mobility.
The joint distribution of the conditioning variables is easily visualized in a HDDS table by the areas of the half-disks. Therefore, when the areas of the HDDSs on the diagonal of the inner part of Figure~\ref{fig:share_HDDS} are bigger than the other cells, this signals low social mobility.
The visual inspection of the inner part of the HDDS table shows that people coming from a poor family are more likely to have low income, while the people born in a rich or average family are more uniformly distributed across income classes.
Notice that this is just a suggestion coming from the visualization of the data with HDDSs, and requires to be supported by formal statistical analyses in order to test the existence of such relationship.

%

\section{Conclusions}   \label{sec:conclusions}

We have introduced the half-disk density strip (HDDS) to visualize probability distributions. This method generalizes the density strips of \cite{jackson2008displaying} and conveys multiple layers of information in an intuitive manner by exploiting color shading and half-disk area size, which facilitates its interpretability also for users without a statistical background.

By arranging a set of HDDSs in an array, one obtains an HDDS table, which extends the concept of contingency table and can be effectively used to visualise a variable of interest, $z$, marginally and conditionally on the  pair $(x,y)$, where all random variables can be categorical, discrete or continuous.
In an HDDS table the distribution of $z$ is represented by the color shading, while the joint and marginal distributions of the conditioning variables, $(x,y)$, is proportional to the area of the HDDSs in each cell of the table.

Moreover, we have shown how HDDSs (and tables) can be efficiently combined to visually compare distributions from different sources, e.g., over time and across countries.
These methods have been used to analyse two real datasets, illustrating the numerous advantages of HDDSs in displaying and comparing distributions, as opposed to standard tools, such as kernel density plots.

\bibliographystyle{plain}
\bibliography{biblio}

\begin{thebibliography}{10}

\bibitem{bertin2011graphics}
Jacques Bertin.
\newblock {\em Graphics and graphic information processing}.
\newblock Walter de Gruyter, 2011.

\bibitem{bertrand2015gender}
Marianne Bertrand, Emir Kamenica, and Jessica Pan.
\newblock Gender identity and relative income within households.
\newblock {\em The Quarterly Journal of Economics}, 130(2):571--614, 2015.

\bibitem{bowman2019graphics}
Adrian~W Bowman.
\newblock Graphics for uncertainty.
\newblock {\em Journal of the Royal Statistical Society: Series A (Statistics
  in Society)}, 182(2):403--418, 2019.

\bibitem{boyce2010money}
Christopher~J Boyce, Gordon~DA Brown, and Simon~C Moore.
\newblock Money and happiness: Rank of income, not income, affects life
  satisfaction.
\newblock {\em Psychological science}, 21(4):471--475, 2010.

\bibitem{chambers2018graphical}
John~M Chambers.
\newblock {\em Graphical methods for data analysis}.
\newblock CRC Press, 2018.

\bibitem{chen2007handbook}
Chun-houh Chen, Wolfgang~Karl H{\"a}rdle, and Antony Unwin.
\newblock {\em Handbook of data visualization}.
\newblock Springer Science \& Business Media, 2007.

\bibitem{cleveland1985elements}
William~S Cleveland.
\newblock {\em The elements of graphing data}.
\newblock Wadsworth Publ. Co., 1985.

\bibitem{fagerland2017statistical}
Morten Fagerland, Stian Lydersen, and Petter Laake.
\newblock {\em Statistical analysis of contingency tables}.
\newblock CRC press, 2017.

\bibitem{Fox2011changing}
Peter Fox and James Hendler.
\newblock Changing the equation on scientific data visualization.
\newblock {\em Science}, 331(6018):705--708, 2011.

\bibitem{gelman2002let}
Andrew Gelman, Cristian Pasarica, and Rahul Dodhia.
\newblock Let's practice what we preach: turning tables into graphs.
\newblock {\em The American Statistician}, 56(2):121--130, 2002.

\bibitem{gordon2015statistician}
Ian Gordon and Sue Finch.
\newblock Statistician heal thyself: have we lost the plot?
\newblock {\em Journal of Computational and Graphical Statistics},
  24(4):1210--1229, 2015.

\bibitem{ihaka2003colour}
Ross Ihaka.
\newblock Colour for presentation graphics.
\newblock In K.~Hornik, F.~Leisch, and A.~Zeileis, editors, {\em Proceedings of
  the 3rd International Workshop on Distributed Statistical Computing}, 2003.

\bibitem{jackson2008displaying}
Christopher~H Jackson.
\newblock Displaying uncertainty with shading.
\newblock {\em The American Statistician}, 62(4):340--347, 2008.

\bibitem{poynton2012digital}
Charles Poynton.
\newblock {\em Digital video and HD: Algorithms and Interfaces}.
\newblock Elsevier, 2012.

\bibitem{seide2020utilizing}
Svenja~E Seide, Katrin Jensen, and Meinhard Kieser.
\newblock Utilizing radar graphs in the visualization of simulation and
  estimation results in network meta-analysis.
\newblock {\em Research Synthesis Methods}, pages 1--10, 2020.

\bibitem{Shneiderman2014big}
Ben Shneiderman.
\newblock The big picture for big data: Visualization.
\newblock {\em Science}, 343(6172):730--730, 2014.

\bibitem{tuckey1977exploratory}
JW~Tuckey.
\newblock {\em Exploratory Data Analysis. 1977}.
\newblock Mass: Addison-Wesley Publishing Company Reading, 1977.

\bibitem{tufte2001visual}
Edward~R Tufte.
\newblock {\em The visual display of quantitative information}, volume~2.
\newblock Graphics press Cheshire, CT, 2001.

\bibitem{vandemeulebroecke2019effective}
Marc Vandemeulebroecke, Mark Baillie, Alison Margolskee, and Baldur Magnusson.
\newblock Effective visual communication for the quantitative scientist.
\newblock {\em CPT: pharmacometrics \& systems pharmacology}, 8(10):705--719,
  2019.

\bibitem{zeileis2009escaping}
Achim Zeileis, Kurt Hornik, and Paul Murrell.
\newblock Escaping rgbland: Selecting colors for statistical graphics.
\newblock {\em Computational Statistics \& Data Analysis}, 53(9):3259--3270,
  2009.

\end{thebibliography}

\appendix
\section{Data description}      \label{sec:apdx_data}

\paragraph{Case study I}
The data come from the Survey on Households, Income and Wealth\footnote{\url{https://www.bancaditalia.it/statistiche/tematiche/indagini-famiglie-imprese/bilanci-famiglie/index.html?com.dotmarketing.htmlpage.language=1}} (SHIW) from the Bank of Italy.
The variable of interest is the net current income (Y), $z$, which is a continuous variable on the  positive real line. The conditioning variables are gender (SEX), $x$, and employment status (QUAL), $y$.
The latter is a categorical variable which in the original dataset can take 7 values: (i) blue-collar, (ii) office worker or school teacher, (iii) cadre or manager, (iv) sole proprietor/member of the arts or professions, (v) other self-employed, (vi) pensioner, (vii) other not-employed. In the empirical application, we have removed all the observations referring unemployed  or retired people and have merged category (iv) and (v), thus ending up with four categories.

This dataset also contains a categorical variable identifying the geographical area of residence of the individuals (AREA3). In the second application, we compare the distribution of income between Northern regions (Piemonte, Valle d'Aosta, Lombardia, Trentino-Alto Adige, Veneto, Friuli-Venezia Giulia, Liguria, Emilia-Romagna) and Southern regions (Abruzzo, Molise, Campania, Puglia, Basilicata, Calabria, Sicilia, Sardegna).

\paragraph{Case study II}
The data has been downloaded from the Survey of Health, Ageing and Retirement in Europe\footnote{\url{http://www.share-project.org/home0.html}} (SHARE wave 7 - 2019).
The self-reported life satisfaction (AC012) is taken as the variable of interest, $z$. It is a discrete variable, taking integer values in $[0,10]$.
As conditioning variables we have considered the current income (RE029), $x$, and the economic status of the family of origin (CC733), that is when the individual was between 0-16 years old, $y$.
The former is a positive continuous variable, hence we have partitioned the positive real line in three intervals: $[0,q_{33})$, $[q_{33},q_{67})$, $[q_{67},q_{100}]$, where $q_{i}$ is the $i$-th percentile. This serves to proxy low, medium and high income levels.
Instead, $y$ is categorical with three values corresponding to poor, average and rich family.


\section{Code}      \label{sec:apdx_code}
The \R package \texttt{hdds} is made publicly available at \url{https://github.com/matteoiacopini/hdds} for reproducing the results of the paper.
It contains the functions for constructing single HDDS and for comparing two HDDSs in a single plot, as well as the code for generating HDDSs tables.
A simplified version in \MATLAB language, which allows to plot simple HDDSs, is also available at the same address.

\end{document}